# Recent Developments in MAS DNP-NMR of Materials

Andrew Rankin[a], Julien Trébosc[a], Frédérique Pourpoint[a], Jean-Paul Amoureux[a,b], Olivier Lafon[a,c]*

[a] *Univ. Lille, CNRS, Centrale Lille, ENSCL, Univ. Artois, UMR 8181 – UCCS – Unité de Catalyse et Chimie du Solide, F-59000 Lille, France*

[b] *Bruker Biospin, 34 rue de l'industrie, F-67166, Wissembourg, France*

[c] *Institut Universitaire de France, 1 rue Descartes, F-75231, Paris, France*

*Corresponding author: olivier.lafon@univ-lille.fr

**Abstract**

Solid-state NMR spectroscopy is a powerful technique for the characterization of the atomic-level structure and dynamics of materials. Nevertheless, the use of this technique is often limited by its lack of sensitivity, which can prevent the observation of surfaces, defects or insensitive isotopes. Dynamic Nuclear Polarization (DNP) has been shown to improve by one to three orders of magnitude the sensitivity of NMR experiments on materials under Magic-Angle Spinning (MAS), at static magnetic field $B_0 \geq 5$ T, conditions allowing for the acquisition of high-resolution spectra. The field of DNP-NMR spectroscopy of materials has undergone a rapid development in the last ten years, spurred notably by the availability of commercial DNP-NMR systems. We provide here an in-depth overview of MAS DNP-NMR studies of materials at high $B_0$ field. After a historical perspective of DNP of materials, we describe the DNP transfers under MAS, the transport of polarization by spin diffusion and the various contributions to the overall sensitivity of DNP-NMR experiments. We discuss the design of tailored polarizing agents and the sample preparation in the case of materials. We present the DNP-NMR hardware and the influence of key experimental parameters, such as microwave power, magnetic field, temperature and MAS frequency. We give an overview of the isotopes, which have been detected by this technique, and the NMR methods, which have been combined with DNP. Finally, we show how MAS DNP-NMR has been applied to gain new insights into the structure of organic, hybrid and inorganic materials with applications in fields, such as health, energy, catalysis, optoelectronics etc.

## 1. Introduction

Solid-state NMR spectroscopy provides unique information on the atomic-level structure and dynamics of materials [1], employed for various technological ends, such as energy, health, mobility, catalysis and construction. As a local characterization technique capable of providing atomic resolution, NMR is especially suitable for the study of amorphous, heterogeneous or disordered materials, including oxide glasses [2], heterogeneous catalysts [3–5] or battery materials [6].

Nevertheless, a major limitation of NMR spectroscopy is its lack of sensitivity. Such low sensitivity is due to the small magnetic moments of nuclear spins, which results in small nuclear magnetizations at thermal equilibrium and slow longitudinal relaxation, two phenomena yielding weak NMR signals. The low sensitivity of NMR limits the observation of (i) surfaces and interfaces, (ii) low-volume samples, such as thin-films or cultural heritage samples, (iii) defects, which control numerous properties of materials, such as the reactivity or the ionic conductivity, (iv) isotopes with low gyromagnetic ratio, γ, or low natural abundance, such as $^{15}$N, $^{17}$O, $^{2}$H, $^{89}$Y, etc, or subject to large anisotropic interactions, such as $^{119}$Sn, $^{195}$Pt or $^{35}$Cl.

Therefore, numerous methods have been proposed to enhance the sensitivity of solid-state NMR spectroscopy. These methods use two complementary routes: (i) the design of more sensitive detection schemes and (ii) the enhancement of the nuclear polarization. The sensitivity of the NMR detection has been significantly increased by the introduction of Fourier Transform NMR spectroscopy [7]. More recently, the sensitivity per spin of NMR detection for solids has also been increased using techniques, such as Magic-Angle Spinning (MAS) NMR probes with cryogenic detection systems, also termed cryoprobe [8], microcoils [9] and non-uniform sampling of the indirect dimensions of multidimensional NMR experiments [10].

The second route involves increasing the nuclear polarization. Such enhancement can be achieved by using high static magnetic fields, $B_0$ [11–13]. Solid-state NMR experiments up to 26 T have been reported using magnets made of low-temperature superconducting (LTS) outer coils in series with high-temperature superconducting (HTS) inner coils. Solid-state NMR experiments at static $B_0$ fields up to 40 T have been carried out using LTS outer coils in series with resistive inner coils [11,13]. Nevertheless, the high running costs of these resistive magnets limit their use. The nuclear polarization can also be enhanced by polarization transfer, such as cross-polarization, from high-γ isotopes to low-γ ones [14,15]. Such polarization transfers are also employed in the indirect detection via high-γ spins, such as protons [16]. Other strategies to increase the nuclear polarization under MAS include low temperature, which has been decreased down to 5 K [17–19], Dynamic Nuclear Polarization (DNP), which consists of a transfer of polarization from unpaired electrons to the nuclear spins [20–24], xenon-129 gas hyperpolarized by Spin Exchange Optical Pumping [25,26], para-hydrogen [27,28] and photochemically-induced DNP [29–31].

Among the approaches listed above, DNP at $B_0 \geq 5$ T under MAS conditions offers several advantages: (i) it can be applied for a wide range of systems, including small organic molecules, biomolecules, organic, hybrid and inorganic materials, (ii) it yields sensitivity gains of several orders of magnitude (10-10$^3$) whereas a doubling of the $B_0$ field strength only improves the sensitivity by a factor of 2.8 for spin-1/2 nuclei, (iii) it is compatible with the acquisition of high-resolution NMR spectra, (iv) it allows the acquisition of multidimensional NMR experiments and (v) companies supply commercial DNP-NMR spectrometers.

We focus here on DNP-NMR of materials at $B_0 \geq 5$ T under MAS conditions. The DNP of materials is as old as the DNP technique itself since the DNP phenomenon was first experimentally demonstrated on lithium metal in 1953 by Carver and Slichter (see Figure 1a) [32]. However, these first experiments were carried out at $B_0 = 3$ mT under static conditions. In 1958, Abragam and Proctor reported the first DNP experiments on a dielectric inorganic material, which was a single crystal of LiF doped with F-centers [33]. DNP was also demonstrated for other dielectric materials, including inorganic single crystals doped with paramagnetic ions, such as $Nd^{3+}$, $Tm^{2+}$ or $Ce^{3+}$ or semi-conductors, such as *n*-doped silicon [34].

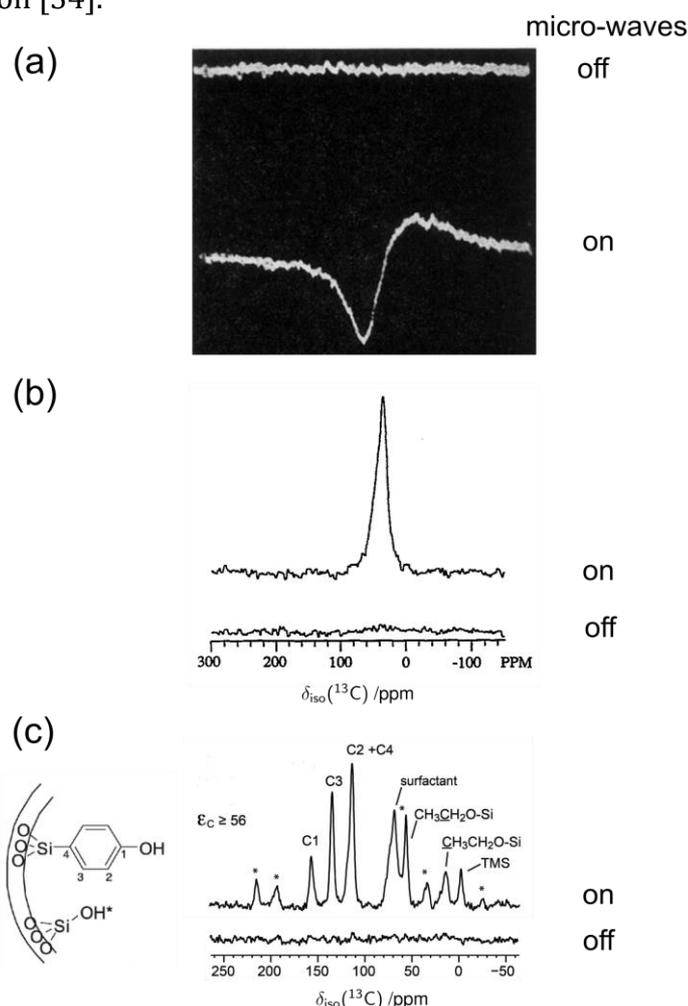

**Figure 1.** Some milestones in the development of DNP-NMR of materials: (a) $^7$Li NMR spectra of lithium metal without (top) and with (bottom) microwave irradiation at 3 mT under static conditions; (b) $^1H \rightarrow ^{13}C$ NMR spectra of CVD diamond film with (top) and without (bottom) microwave irradiation at 1.4 T and room temperature with a MAS frequency $\nu_R$ = 3.8 kHz at room temperature; (c) $^1H \rightarrow ^{13}C$ CPMAS NMR spectra of mesoporous silica functionalized with phenol moieties impregnated with TOTAPOL aqueous solution with (top) and without (bottom) microwave irradiation at 9.4 T with $\nu_R$ = 8 kHz. The numbering of the carbon atoms of the phenol moieties is displayed on the left. Adapted with permission from refs. [32,35,36]. Copyright 1953, American Physical Society, 1993, AIP Publishing and 2010, American Chemical Society.

In 1983, Wind reported the first DNP-NMR experiments under MAS conditions [20]. The MAS technique allowed for the acquisition of high-resolution DNP-enhanced NMR spectra of materials, such as coal, diamonds, polymers and organic conductors [22]. The sensitivity gain provided by DNP was notably used to probe the interface of an immiscible mixture of polycarbonate and polystyrene [37] and the surface of chemical vapor deposited (CVD) diamond film (see Figure 1b) [35,38]. However, these MAS DNP-NMR experiments were constrained to $B_0 \leq 1.5$ T, i.e. $^1$H Larmor frequency $\nu_0(^1H) \leq 60$ MHz because of the paucity of microwave sources operating above 40-50 GHz.

DNP-enhanced NMR at $B_0 \geq 5$ T under MAS conditions was pioneered by the group of Griffin at MIT in the 1990s. Major developments included (i) the introduction of cyclotron resonance masers, known as gyrotrons, into DNP experiments as a continuous high-power microwave source at frequencies higher than 140 GHz [39], (ii) the design of cryogenic MAS probes that operate at temperatures of 90 K and below [21,40], and (iii) the design of nitroxide biradicals, which generally yield more efficient DNP transfer at high fields than monoradicals [41]. The MAS DNP spectroscopy at high fields was initially mainly applied to the study of solid-state biomolecules and notably allowed for the observation of photocycle intermediates of bacteriorhodopsin, a proton pump of Archea [42].

The availability of a commercial MAS DNP-NMR system [43] at $B_0$ = 9.4 T has led to the use of this technique for the characterization of materials. The possibility of applying high-field MAS DNP-NMR to materials was first reported by the groups of Emsley, Copéret and Bodenhausen in 2010 [36]. They showed that DNP at 9.4 T could yield a 50-fold enhancement of $^1H \rightarrow {}^{13}C$ Cross-Polarization under MAS (CPMAS) signals of phenol moieties covalently bonded to the surface of mesoporous silica impregnated with an aqueous solution of nitroxide biradicals (see Figure 1c). Independently, our group demonstrated that DNP can yield a 30-fold enhancement of $^{29}Si$ signals in direct excitation experiments under MAS of mesoporous silica impregnated with a solution of nitroxide biradicals in a DMSO/water mixture [44,45]. Owing to the availability of commercial MAS DNP-NMR systems, this technique is applied for the characterization of a rapidly increasing number of organic, hybrid and inorganic materials (see Figure 2).

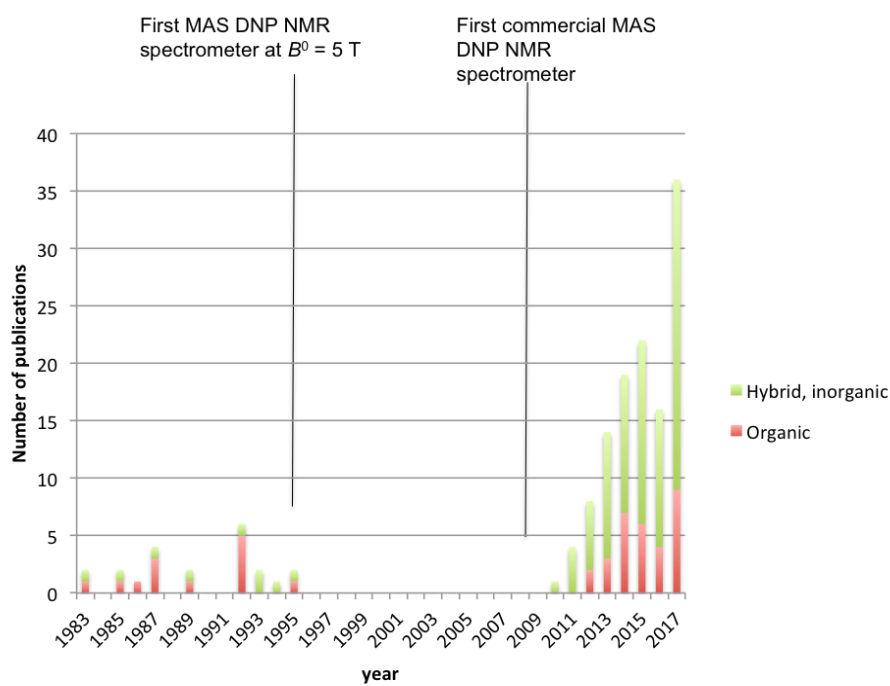

**Figure 2.** Number of publications per year relating to MAS DNP-NMR spectroscopy of organic (red) and hybrid or inorganic (green) materials since the introduction of this technique.

We review herein the recent publications relating to high-field MAS DNP-NMR experiments on materials since 2010. Hence, we mainly restrict ourselves to articles reporting on materials with this technique, even if some seminal results about MAS DNP-NMR of small molecules or biomolecules in frozen solutions are also mentioned. The interested reader is referred to other recent review articles on that topic [4,23,24,46–65]. We will first describe the transfers of polarization in MAS DNP and the various factors contributing to the global sensitivity of MAS DNP-NMR experiments. We then present a selection of the various polarizing agents (PAs), which have been employed for high-field MAS DNP-NMR, and the preparation of material samples for these experiments. We describe the MAS DNP-NMR spectrometers used and the influence of various experimental conditions ($B_0$ field, microwave power, sample temperature, MAS frequency…) on the DNP enhancement. We present the various isotopes and NMR experiments, for which MAS DNP has been employed in the case of materials. Finally, we provide an overview of applications of high-field MAS DNP for the characterization of organic, hybrid and inorganic materials.

## 2. Polarization transfer and global sensitivity

### 2.1. DNP mechanisms and depolarization

In materials, MAS DNP transfers at $B_0 \geq 5$ T have been mainly achieved using the solid effect (SE), cross effect (CE) and Overhauser effect (OE). The thermal mixing is another DNP mechanism, which has been reported in the case of many strongly coupled unpaired electrons exhibiting homogeneously broadened EPR line [66–68]. However, the EPR lines are generally inhomogeneously broadened during MAS DNP NMR experiments and hence, the thermal mixing will not be discussed further [52]. The measurement of the DNP enhancement as function of the $B_0$ field, i.e. the DNP field profile, is often useful to identify the involved DNP mechanisms [50,53].

The SE mechanism involves an unpaired electron and a nucleus. It dominates when the width of the EPR lines is much narrower than the Larmor frequency of the polarized isotope, $\nu_0(I)$. Hence, it has been observed for unpaired electrons with symmetric local environments, such as 1,3-bisdi-phenylene-2-phenylallyl (BDPA) [69], as well as Mn(II), Gd(III) and Cr(III) ions [59,70–73]. The SE relies on the microwave irradiation of the forbidden zero-quantum or double-quantum transitions at frequencies of $\nu_0(I) \pm \nu_0(S)$, where $\nu_0(S)$ denotes the Larmor frequency of the unpaired electron. A difference under MAS with respect to static conditions is that the frequencies of transitions vary with the rotation of the sample, instead of being time-independent. Hence, these transitions under MAS are irradiated periodically, instead of continuously, as is the case under static conditions [52,74,75]. Under MAS, the irradiation of these forbidden transitions can be viewed as adiabatic level crossings and hence, the polarization transfers between the unpaired electron and the nucleus are less efficient at higher $B_0$ field and MAS frequency. Such decrease can be counteracted by increasing the microwave power.

The CE mechanism is often the most efficient one under the conditions of high-field MAS DNP. It involves two unpaired electrons, $S_a$ and $S_b$, coupled by $J$-exchange and/or dipolar interactions and a nucleus, $I$. It is notably observed when the **g**-tensor anisotropy of at least one of the two unpaired electrons dominates the EPR linewidth and exceeds the Larmor frequency of the polarized nucleus, $\nu_0(I)$. It has been shown that the CE mechanism under MAS differs from that under static conditions [52,74–76]. Under MAS, during one rotor period, the instantaneous resonance frequency of an unpaired electron, $\nu_0(S_i)$ with $i = a$ or $b$, moves across the EPR line, which leads to several crossings and anti-crossings of energy levels (see Figure 3). These crossings and anti-crossings can produce population exchanges between the energy levels, provided they are long enough, i.e., adiabatic [74]. During the rotor period, when the microwave irradiation is resonant with one of the two unpaired electrons, i.e., its instantaneous EPR frequency is equal to the carrier frequency of microwaves, this irradiation can create a difference in polarization between the coupled unpaired electrons, which exceeds the nuclear polarization at thermal equilibrium. Such a difference can only be created if the microwaves irradiate one electron, without perturbing the other one. Strong exchange and dipolar couplings between the unpaired electrons lead to an exchange of polarization between them and allow the polarization difference between them to be maintained during the electron-electron anti-crossings. For instance, in Figure 3, the electrons *b* and *a* are irradiated by microwave during the first and the second halves of the rotor period, respectively. The efficient polarization exchange between them at the electron-electron dipolar anti-crossing in the middle of the rotor period allows the polarization difference between the two electrons not to be reduced by the irradiation of the electron $a$ during the second half of the rotor period. The polarization difference between the unpaired electrons is periodically transferred to the nucleus during CE rotor events, occurring when $|\nu_0(S_a) - \nu_0(S_b)| \approx |\nu_0(I)|$. Unlike under static conditions, the microwave crossings and CE anti-crossings do not have to occur simultaneously.

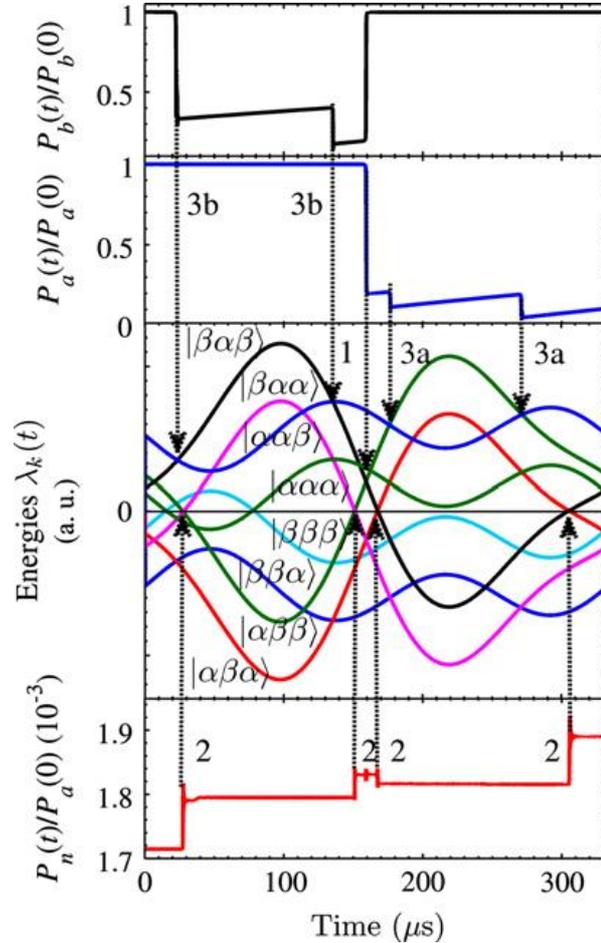

**Figure 3.** Evolution of the polarizations of unpaired electrons, $P_i(t)$ with $i = a$ or $b$ (top), the energy levels (middle) and the nuclear polarization, $P_n(t)$ (bottom), of a spin system containing two coupled unpaired electrons and a nucleus during a steady-state rotor period at $B_0$ = 9.4 T, a temperature of 100 K with $\nu_R$ = 3 kHz. The rotor-events marked on the figure are (1) the electron-electron dipolar anti-crossing, (2) the CE anti-crossings and (3) the microwave crossings. Adapted with permission from ref. [75]. Copyright 2012, Elsevier.

The difference in polarization between the unpaired electrons decreases for increasing MAS frequency and $B_0$ field since these parameters reduce the adiabaticity of crossings and anti-crossings [76]. Nevertheless, the decrease of DNP enhancement at higher fields is lower for MAS CE than for MAS SE [76]. For MAS CE, the polarization difference between unpaired electrons can generally be enhanced by increasing the microwave power and the strength of the exchange and dipolar couplings between electrons. However, a saturation or even a decrease of the DNP enhancement is observed for very large microwave power [43,56,77,78]. Furthermore, larger exchange and dipolar interactions between unpaired electrons lead to more efficient CE anti-crossings, which accelerate the build-up of the nuclear polarization [74,79,80]. Nevertheless, CE transfers are only efficient provided the electron-electron exchange and dipolar interactions remain smaller than $\nu_0(I)$ [81]. In addition, a larger microwave intensity is needed in the case of shorter electron longitudinal and transverse relaxation times in order to obtain the same difference in the polarization of unpaired electrons [74,76]. Numerical simulations of $^1$H CE at 9.4 T and 100 K with a MAS frequency of 8 kHz have shown that a maximal DNP enhancement using nitroxide biradicals as PAs is achieved when (i) the $g_{yy}$ tensor components of both electrons are perpendicular to one another, (ii) the number of covalent bonds linking the two nitroxide moieties is greater than or equal to three, and (iii) the electron longitudinal relaxation times $T_1(S_i)$ are near 500 µs [82].

Furthermore, it has been shown that MAS-DNP simulations are accurate enough to reproduce the field profile for DNP of protons using bis-nitroxide PAs (see Figure 4) employing as input parameters (i) the g-tensors, (ii) their relative orientation, (iii) the electron-electron dipolar coupling, (iv) the *J*-exchange interaction, (v) the hyperfine coupling and (vi) the electron relaxation times, which were determined using DFT calculations and high-field EPR measurements, as well as (vii) the nuclear relaxation times, (viii) the MAS frequency and (ix) the microwave frequency and amplitude [83]. The ratio between the positive maximum and the negative minimum of the DNP field profile can only be reproduced correctly, when taking into account the dependence of the electron longitudinal relaxation time with the PA orientation [83,84]. Conversely the nuclear relaxation times and the electron-proton hyperfine coupling values have, within reasonable limits, little effect on the shape of the DNP field profile, but they strongly affect the absolute DNP enhancement. Simulations of the build-up of DNP-enhanced $^1$H magnetization have been used to estimate the longitudinal relaxation time of protons close to the PA.

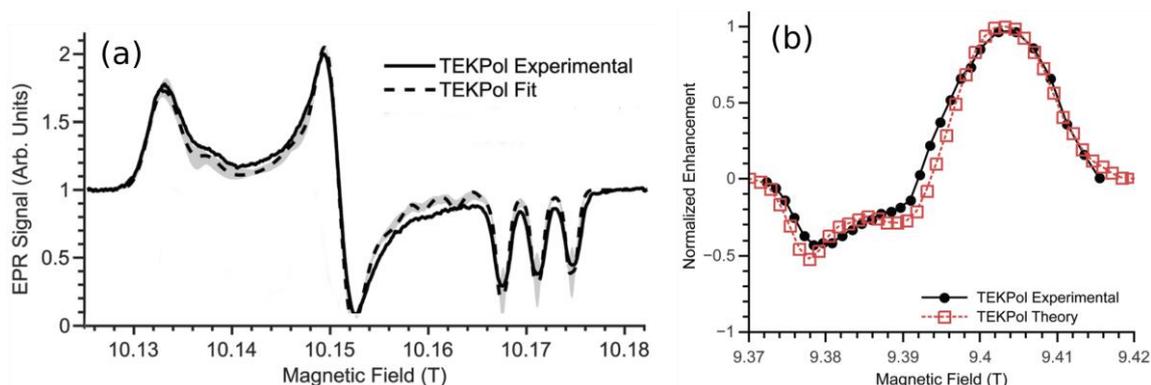

**Figure 4.** (a) Experimental EPR spectrum (solid line) of 15 mM frozen solution of TEKPol PA (see **Figure 6**) in a mixture of CHCl$_3$/1,1,1,2-tetrabromoethane/[$^2$H$_4$]-methanol at 100 K with a microwave frequency of 285 GHz. The best-fit simulation (dashed line) of the experimental spectrum is also displayed. The grey overlay corresponds to the error bar for the best-fit simulation. (b) Experimental (black circles) and simulated (red squares) field profile for MAS DNP of protons in the same sample as in subfigure a at 9.4 T and 110 K with $\nu_R$ = 8 kHz. Adapted with permission from ref. [83]. Copyright 2019, Royal Society of Chemistry.

Long nuclear longitudinal relaxation time, $T_1(I)$, is needed so that the steady state nuclear polarization can equal the difference of polarization between the unpaired electrons. As a consequence, the presence in the materials of paramagnetic centers, which do not contribute to the DNP transfer but shorten the electron and nuclear longitudinal relaxation times, $T_1(S_i)$ and $T_1(I)$, respectively, decreases or even cancels the DNP enhancement. This polarization leakage phenomenon has been reported for materials containing endogeneous paramagnetic centers, such as natural clays and functional nanodiamonds [85,86]. Similarly, it has been shown that the removal of the paramagnetic molecular oxygen (O$_2$) dissolved in organic solvents or adsorbed in polymers can improve the DNP enhancement [87,88]. It has also been demonstrated that the chemical passivation of the surface of mesoporous silica by methyl or $OSi(CH_3)_3$ groups strongly reduces the $T_1(^1H)$ value and the DNP enhancement [89]. The replacement of the fast-relaxing protons by deuterons however mitigates this effect.

In the absence of microwave irradiation, the CE anti-crossings can also reduce the nuclear polarization, which can then be lower than its value at thermal equilibrium. This depolarization phenomenon has been demonstrated theoretically and experimentally [90,91]. It has been shown that the magnitude of the depolarization depends on the PA. In particular, longer $T_1(S_i)$ times lead to stronger nuclear depolarization. The depolarization does not affect the sensitivity of DNP NMR experiments. However, a consequence of this effect is that the DNP enhancement factor, $\varepsilon_{on/off}$, estimated as the ratio between signal intensities with and without microwave irradiation, does not provide an accurate measure of the polarization enhancement with respect to thermal equilibrium.

Another DNP mechanism, which has been employed for high-field MAS DNP of materials, is the Overhauser effect (OE) [69,92–95]. In the OE, the EPR transition is saturated by microwave irradiation, while the stochastic modulation of the electron-nuclear interactions leads to cross-relaxation, which enhances the nuclear polarization [24]. Recently a positive OE has been reported in materials containing 1,3-bisdi-phenylene-2-phenylallyl (BDPA) radicals [93]. Furthermore, the DNP enhancement of protons in those materials increases with increasing $B_0$ strength. Hence, the OE has been mainly applied at $B_0$ = 18.8 T [69,94,95]. The origin of the stochastic modulation of the electron-nuclear interactions for BDPA radicals incorporated into solids is still under debate [96,97].

## 2.2. Spin diffusion

The enhancement of nuclear polarization by DNP also often relies on spin diffusion, which transfers the enhanced polarization of *I*-spin nuclei in the vicinity of the PAs to the remote identical nuclei through the homonuclear dipolar interactions. It has been shown that *I* nuclei subject to hyperfine interactions can exchange polarization under MAS during homonuclear dipolar rotor events, where their resonance frequencies are equal [76,79]. This model suggests that the spin diffusion barrier under MAS is porous, contrary to the static case [98,99]. The characteristic length of spin diffusion is equal to

$$l_D = \sqrt{DT_1(I)} \qquad (1)$$

where $D$ is the spin diffusion coefficient [100–103]. This coefficient increases with increasing homonuclear dipolar coupling between *I* nuclei. Hence, $l_D$ length increases with larger homonuclear dipolar coupling and slower longitudinal relaxation. For instance, spin diffusion is often more efficient at low temperature, which reduces motions at atomic level. For protons in organic materials, the $l_D$ values typically range from 10 nm to 1 μm at about 100 K, depending on the $T_1(^1H)$ value, which can reach $10^3$ s for organic crystals of molecules, which do not possess methyl groups [103].

$^1H$ spin diffusion is always involved in experiments using DNP-enhanced $^1H$ polarization. Nevertheless, it is especially useful for materials, for which the PAs cannot be brought within a few nanometers of the nuclei to polarize. For instance, $^1H$ spin diffusion has been employed in microcrystalline solids, including pharmaceuticals, to transport DNP-enhanced polarization from the surface into the interior of the crystals [65,101,103,104]. $^1H$ spin diffusion transport has also been applied to non-porous inorganic solids, such as magnesium and calcium hydroxide [105], hydroxyapatite [106,107], ammonia borane [108] or proton conductors [109]. It has also been employed to study porous materials, including mesoporous silica [102,110], Metal-

Organic Frameworks (MOFs) [111–113] and zeolites [114–117], for which the PAs are too large to diffuse into the pores. The exclusion of the PAs from the pore can prevent their reaction with the catalytic active sites located within the pores [110,114–117]. In the case of aggregated silica nanoparticles functionalized with reactive species, it has also been shown that spin diffusion transports DNP-enhanced $^1$H polarization in the interparticle spaces, whereas PAs are excluded from those spaces and hence are not in contact with reactive surface species [118]. DNP measurements combined with spin diffusion models have also been used to determine the domain sizes in materials, such as organic microcrystals, mesoporous silica and polymer coatings (see Figure 5) [103,104]. Recently, it has also been shown that spin diffusion between slow relaxing spin-1/2 nuclei, other than protons, such as $^{29}$Si, $^{119}$Sn and $^{113}$Cd, at natural abundance contributes to transport the DNP-enhanced polarization from the surface to the interior of inorganic microcrystals [119].

When the size, $L$, of the domain polarized by a PA is small with respect to $l_D$ length, spin diffusion is fast enough to equalize the polarization of all nuclei in the domain. A mono-exponential build-up is then observed ($\beta$ = 1 in Eq. 2) and the polarization buildup time constant, $T_B$, is identical with and without microwave irradiation (see Figure 5b) [79,101,103]. Hence, as seen in Figure 5c, the DNP enhancement is independent of the polarization delay. Conversely, when $L \gg l_D$, the polarization of nuclei closer to the PAs increases faster than that of the remote ones. In the case of particles with PAs on their surface (see Figure 5d), this effect results in faster polarization build-up with microwave irradiation than without [85,119]. As a consequence, the polarization build-up cannot be fitted using a single exponential function but follows a stretched exponential build-up of the form:

$$I = I_\infty \left\{1 - \exp\left[-\left(\frac{t}{T_B}\right)^\beta\right]\right\} \quad (2)$$

with $\beta$ < 1, and where $I_\infty$ is the asymptotic intensity for $t \gg T_B$. Furthermore, the $T_B$ time constant increases with increasing $L/l_D$ ratio and the DNP enhancement depends on the polarization delay (see Figure 5e) [103].

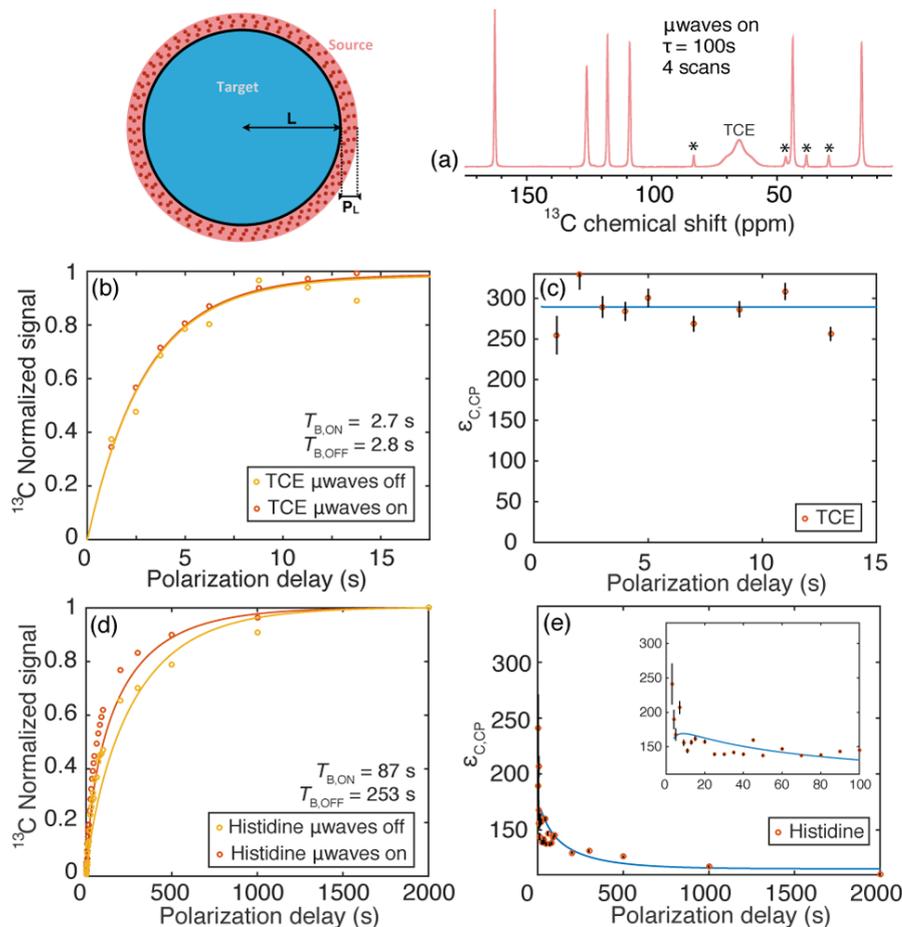

**Figure 5.** $^1H\rightarrow{}^{13}C$ CPMAS experiments of a microcrystalline powder of L-histidine hydrochloride monohydrate impregnated with 16 mM TEKPol (see **Figure 6**) in 1,1,2,2-tetrachloroethane (TCE) at 9.4 T and 105 K with $\nu_R$ = 8 kHz. (a) Spectrum with microwave irradiation and polarization delay of 100 s. (b,d) Normalized build-up and (c,e) DNP enhancement of the (b,c) TCE and (d,e) histidine signals as function of the polarization delay. In subfigure e, an inset shows an expansion of the region corresponding to short polarization delays. In subfigures d and e, the best fit curves to the numerical spin diffusion model are displayed as solid lines and corresponds to $L$ = 2.3 µm, which agrees well with the histidine particle size measured by scanning electron microscopy. Figure adapted from ref. [103]. Copyright 2017, American Chemical Society.

### 2.3. Global sensitivity

Besides DNP transfer and spin diffusion, several additional factors affect the sensitivity of MAS DNP NMR experiments. These factors depend on the sample (properties of PAs, preparation of the sample, etc) as well as the DNP NMR system hardware ($B_0$ field, MAS frequency, microwave power in the sample, sample temperature, etc). The unpaired electrons of the PAs also reduce the apparent transverse relaxation of nearby nuclei, leading to line broadening, known as 'paramagnetic quenching', since it prevents the observation of nuclei distant by less than a few Angstroms from the unpaired electrons [91,120–122]. This quenching has been found to be approximately equal to 15% for a concentration of nitroxide biradicals of 12 mM and to be independent of the MAS frequency over an interval of 0 to 10 kHz [91]. In principle, high MAS frequencies would better average the hyperfine interaction and reduce the paramagnetic quenching. However, such effects have not been reported so far. The unpaired electrons also shorten the lifetime of coherences, which further reduces the sensitivity of NMR experiments employing several delays and pulses [123]. Conversely, the unpaired electrons accelerate the build-up of the DNP-enhanced nuclear polarization, allowing for shorter recycle delays and hence, enhancing the sensitivity.

As described in section 3, the DNP-NMR samples are often prepared by impregnating or wetting the investigated material with a solution of radicals. Besides the radicals, the presence of the frozen solvent also alters the sensitivity of DNP-NMR experiments. It can (i) dilute the amount of the material of interest in the rotor, (ii) broaden the NMR lines because of the distribution of local environments in the bulk and at the surface of the frozen solvent, (iii) modify the efficiency of the NMR experiments, such as CPMAS, and (iv) change the nuclear relaxation times [55,121,122,124].

Furthermore, DNP NMR experiments are usually performed at low temperature around 100 K. At such temperature, the density of $N_2$ gas is increased with respect to room temperature and the speed of sound in $N_2$ is significantly reduced. As a consequence, DNP-NMR experiments have to be performed with smaller diameter rotors to obtain the same MAS frequency as for conventional NMR experiments at room temperature. The use of smaller rotors, for instance with a diameter of 3.2 mm instead of 4 mm, results typically in a 30% sensitivity loss.

One approach to estimate the global sensitivity of a DNP-NMR experiment consists of comparing its sensitivity, *i.e.*, its signal-to-noise ratio ($S/N$) per unit of time, with that of conventional NMR experiments [55,60,121–123]. However, this approach is only applicable when NMR signals can be detected using conventional NMR experiments. The global sensitivity gain, $\varepsilon_{\text{global}}^{\text{time}}$, of DNP NMR experiments with respect to conventional NMR experiments can be separated into different factors

$$\varepsilon_{\text{global}}^{\text{time}} = \varepsilon_{\text{on/off}}^{\text{scan}} \varepsilon_{\text{para}}^{\text{scan}} \varepsilon_{\text{solvent}}^{\text{scan}} \varepsilon_{\text{probe}}^{\text{scan}} \varepsilon_{\text{seq}}^{\text{scan}} \varepsilon_{\text{B}}^{\text{scan}} \varepsilon_{\text{T}}^{\text{scan}} \left(\frac{T_1}{T_B}\right)^{1/2} \qquad (3)$$

where $\varepsilon_{\text{on/off}}^{\text{scan}}$ is the DNP enhancement calculated as the ratio between the NMR signals with and without microwave irradiation, $\varepsilon_{\text{para}}^{\text{scan}}$ is the paramagnetic quenching, the factors $\varepsilon_{\text{probe}}^{\text{scan}}$, $\varepsilon_{\text{seq}}^{\text{scan}}$, $\varepsilon_{\text{B}}^{\text{scan}}$ and $\varepsilon_{\text{T}}^{\text{scan}}$ quantify the $S/N$ modifications produced by the presence of frozen solvent and differences in probe, pulse sequence, static magnetic field and sample temperature when comparing DNP and conventional NMR experiments. The last term in Eq. 3 accounts for the difference in the build-up time constants of the longitudinal magnetization of the excited isotope between DNP and conventional NMR experiments. Recently it has also been reported that the DNP-enhanced $^1$H polarization, near the unpaired electrons or averaged over the whole sample, can be quantified by measuring the amplitude of the longitudinal two-spin order, $I_z^H I_z^C$, of coupled $^1$H-$^{13}$C nuclei at the end of the polarization delay [125].

## 3. Polarizing agents and sample preparation

### 3.1. Polarizing agents

Various PAs have been employed for high-field MAS DNP-NMR, including biradicals, BDPA, trityl, paramagnetic ions and endogeneous paramagnetic centers. The structures of some of the radicals used for high-field MAS DNP are displayed in Figure 6. All these PAs exhibit an EPR transition at a *g* factor near 2, since commercially available gyrotrons only deliver microwaves at a fixed frequency, which is typically set tuned to that for nitroxides.

Most of high field MAS DNP-NMR experiments on materials have been carried out using exogeneous nitroxide biradicals. These PAs have been designed to provide a significant MAS CE at high field. In these molecules, the covalent link between the two nitroxide radicals leads to an intramolecular dipolar coupling. Optimal DNP enhancements are obtained with lower concentrations of biradical PAs than monoradical PAs, reducing the paramagnetic quenching. The first nitroxide biradical, which was used for high-field MAS DNP of materials, is TOTAPOL [36,126]. It has been shown that the TEMPO rings of TOTAPOL are approximately perpendicular to one another [127]. This favorable orientation results in many CE anti-crossings during one rotor period, which facilitate the DNP transfer [75]. TOTAPOL is soluble in water as well as in glass-forming water/glycerol and water/DMSO mixtures.

For the study of water-incompatible materials, the first exogeneous nitroxide biradical, which has been used, is bTbk, which is soluble in halogenated solvents [128,129]. In a bTbk molecule, a rigid tether constrains the relative orientation of the TEMPO moieties, which are roughly orthogonal. Such constrained relative orientation favors the CE and leads to DNP enhancements, which are 1.4 times larger than those obtained with TOTAPOL under the same conditions (in DMSO/water solvent) [128]. A further improvement has consisted in slowing the longitudinal and transverse relaxations of the unpaired electrons, which permits to lower microwave power for a given DNP enhancement [76]. The replacement of the methyl groups by a cyclohexyl ring lengthens the transverse relaxation time (also called phase memory time). This approach was used to design the bCTbk radical (see Figure 6) [130], which is soluble in halogenated solvents and leads to $\varepsilon_{on/off}^{scan}$ values, which are 2.5 times larger than those of bTbk. A further refinement was to increase the molecular weight of the biradical in order to lengthen the longitudinal relaxation time of the electron since the relaxation *via* the Raman process, which is the dominant mechanism under DNP conditions used for materials, is less efficient for heavier radicals. This strategy led to the design of TEKPol (see Figure 6) and TEKPol2 [89,131], which are soluble in halogenated solvents and result in sensitivity gains about 3.3 times larger than those of bTbk. A similar strategy was used to design a water-soluble nitroxide biradical, AMUPol, which resulted in DNP enhancements 3.3 times larger than TOTAPOL [132]. Nevertheless, the slower electron relaxation for AMUPol than TOTAPol results in a stronger depolarization, which partially inflates the DNP enhancement of AMUPol [91]. AMUPol and TEKPol are commercially available, which explains their wide use in high-field MAS DNP.

As mentioned in section 2.1, exchange and dipolar interactions between the unpaired electrons, which are large but smaller than $v_0(I)$, improve the efficiency of the CE transfer by (i) maintaining the polarization difference between the two electrons and (ii) improving the efficiency of the CE anti-crossings, which accelerates the build-up of DNP-enhanced nuclear polarization. These theoretical guidelines led to the design of the water-soluble AsymPolPOK nitroxide biradical [80]. In such a radical, the nitroxide moieties are tethered by a short linker, which leads to sizable exchange and dipolar interactions between the unpaired electrons. AMUPol and AsymPolPOK are currently the most efficient water-soluble PAs at 9.4 T and 100 K.

Furthermore, when the PAs are close to the observed nuclei, they can produce paramagnetic quenching and reduce the coherence lifetime, which results in increased losses in NMR experiments. Moreover, the PAs may interact or even chemically react with the surface of the material [133,134]. This limitation has been circumvented by the development of carbosilane-based dendritic nitroxide biradicals, in which the bulky dendrimer attached to the biradical reduces the interaction of the latter with the investigated material [135]. The reaction between reactive surface species and nitroxide biradicals can also be prevented by protecting those species inside pores with entrances smaller than the radical size [110,115] or inside the spaces between aggregated nanoparticles [118] as well as by sequestrating the radicals inside the walls of silica materials [136].

BDPA dissolved in the glass-forming *ortho*-terphenyl (OTP) solvent has been employed for DNP via the SE and the OE [69,94]. For increasing $B_0$ field, the DNP enhancement decreases strongly for SE, whereas it increases slightly for the OE [93]. Nevertheless, the sensitivity enhancement for BDPA is generally smaller than that of TEKPol under similar experimental conditions because the build-up time of the nuclear polarization is by one order of magnitude slower for the OE with BDPA than for the CE with TEKPol.

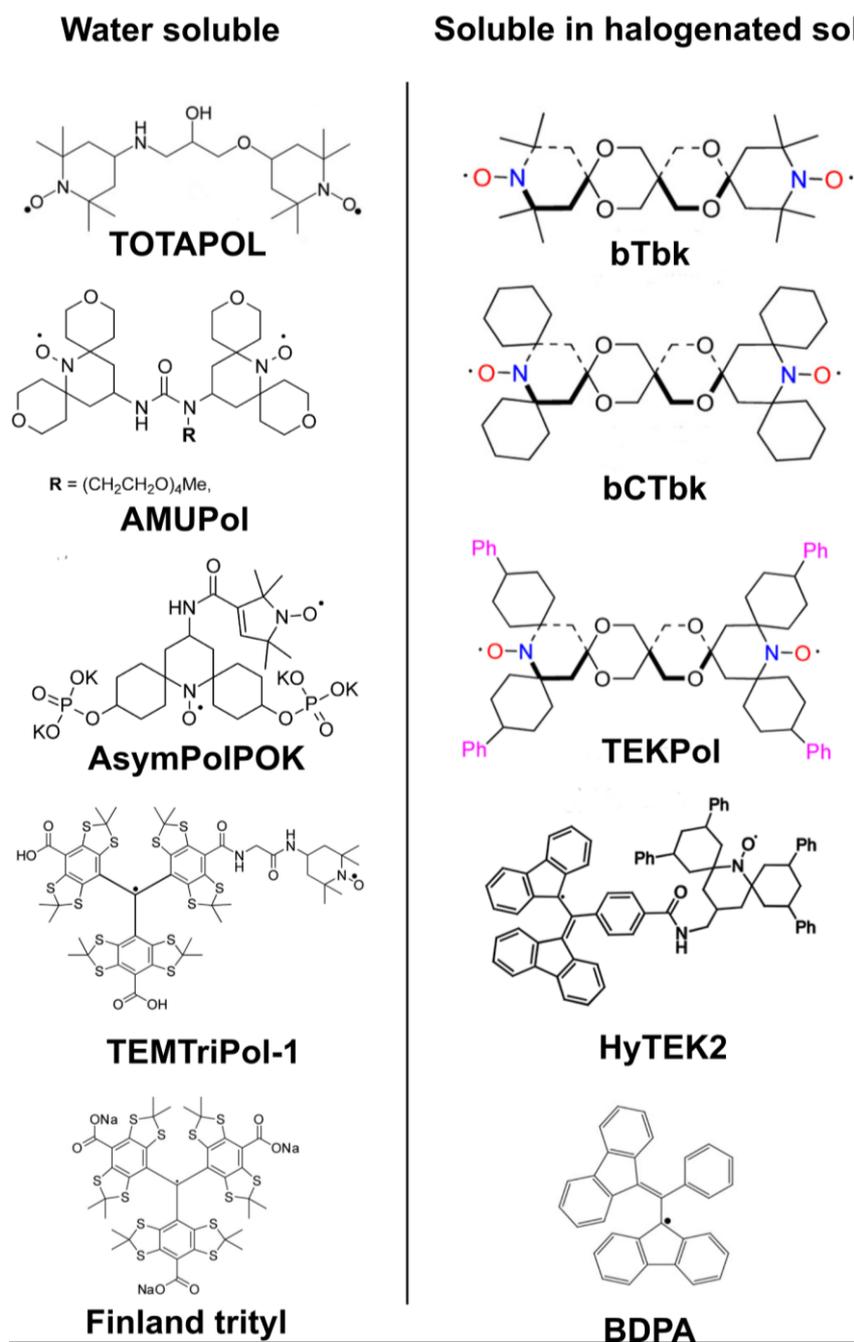

**Figure 6.** Structures and names of some of the organic radicals used as PA for high-field MAS DNP of materials. Figures adapted with permission from refs. **[48,80,132,137–140]**. Copyright 2011, Elsevier, 2013 and 2018 American Chemical Society, 2016, Wiley, 2018 Royal Society of Chemistry,

The DNP enhancement yielded by nitroxide biradicals drastically decreases at higher field. This decrease is related to the broadening of the EPR lines of these biradicals, which are dominated by the **g**-tensor anisotropy and hence, are proportional to the $B_0$ field. Such broadening results in faster frequency sweeps during the crossings and the anti-crossings and hence, to a lower adiabaticity for the population exchange. A strategy to limit the decrease of DNP enhancement at high-field would involve the use of biradicals with narrower EPR lines than nitroxide. Nevertheless, the CE transfer requires that the two electrons of the biradical fulfill the condition $|\nu_0(S_a) - \nu_0(S_b)| \approx |\nu_0(^1H)|$ during the CE rotor-events (see subsection 2.1). This approach has led to the development of the water-soluble TEMTriPol-1 biradical composed of a narrow-line trityl radical tethered to a broad-line TEMPO [81,138]. This narrow-line trityl radical results in more adiabatic crossings and anti-crossings and so requires notably less microwave power. The difference between the isotropic g values of these two unpaired electrons corresponds approximately to the $^1H$ Larmor frequency. Hence, an efficient intramolecular CE transfer can be achieved between the trityl and TEMPO unpaired electrons, when the trityl EPR transition is irradiated. Furthermore, the exchange interaction between its two unpaired electrons can exceed 100 MHz for some of the conformers, which helps to maintain the polarization difference between the unpaired electrons and improves the efficiency of CE anti-crossings [138]. TEMTriPol-1 is currently the most efficient water-soluble PA at 18.8 T and 100 K. More recently, other trityl-nitroxyl biradicals, which are soluble in halogenated solvents, have been prepared and tested for MAS DNP at 9.4 T and 100 K but yielded lower DNP enhancements than TEKPol under those conditions [141].

A similar approach was later applied to develop the HyTEK2 biradical, which is composed of a narrow-line BDPA and a broad-line TEMPO and is soluble in halogenated solvents [139]. Here again, the isotropic g values of the two unpaired electrons differ approximately by the $^1H$ Larmor frequency. Furthermore, BDPA benefits from a slower electron longitudinal relaxation and a narrower EPR line than trityl, and so requires lower microwave power. HyTEK2 also exhibits a significant exchange interaction, $|J| \approx$ 40 MHz, which accelerates the polarization build-up. Furthermore, the nitroxide moiety of HyTEK2 is substituted by cyclohexyl rings with phenyl substituents in order to slow the nitroxide electron relaxation. The HyTEK2 biradical is currently the most efficient PA in halogenated solvents at 18.8 T and 100 K.

The biradicals presented above were all optimized for the CE polarization transfer between unpaired electrons and protons. As the CE condition $|\nu^0(S_a) - \nu^0(S_b)| \approx |\nu^0(I)|$ depends on the Larmor frequency of the polarized nucleus, $I$, PAs with narrower EPR line can be used for DNP of isotopes with lower γ ratios. For instance, it has been shown by the Griffin research group that trityl radicals yield 4-fold larger DNP enhancement for the direct CE of $^2H$ at 5 T and 90 K with $\nu_R$ = 6 kHz than TOTAPOL with the same concentration of unpaired electrons [142]. Similarly, 15-fold larger enhancements were measured for the direct CE of $^{17}O$ using trityl instead of TOTAPOL at 5 T and 90 K under static conditions [143]. Furthermore, the DNP enhancement of the direct CE of $^{13}C$ nuclei at 5 T and 90 K with $\nu_R$ = 4.8 kHz was improved by 30% using an equimolar mixture of trityl and a sulfonated derivative of BDPA, instead of trityl [144]. More recently, significant polarization gains were reported using trityl radical for the direct DNP of $^{13}C$ and $^{15}N$ nuclei at 14.1 T and 100 K with $\nu_R$ = 10 kHz [145]. Nevertheless, at 9.4 T and 100 K with $\nu_R$ = 12.5 kHz, a 4-fold larger sensitivity gain was achieved for the $^{17}O$ CE of mesoporous silica impregnated with a TEKPOL solution in TCE, instead of aqueous Finland trityl solution [140].

A limitation of the above biradical PAs is that they cannot be easily incorporated into the bulk of inorganic materials. For such materials and in the absence of an efficient spin diffusion transport, DNP-NMR using these PAs only enhances the polarization of nuclei near the surface. It has been shown that this limitation can be circumvented by doping inorganic or hybrid materials with paramagnetic metal ions (see Figure 25) [71–73]. Nevertheless, these ions must exhibit a narrow resonance centered at g ≈ 2 so that they can be efficiently irradiated by the microwaves with fixed frequency delivered by the gyrotron. So far only three paramagnetic metal ions have yielded significant enhancement in high-field MAS DNP experiments: Cr(III), Mn(II) and Gd(III) ions, which have electronic configurations $3d^3$, $3d^5$ and $4f^7$, respectively, and ground states with high half-integer-spin $S$ = 3/2, 5/2 and 7/2, respectively. The EPR spectra of Cr(III) and Gd(III) ions display a narrow central transition (CT) between the $m_S$ = ±1/2 energy levels at g ≈ 2 and broad satellite transitions (STs) between energy levels $m_S$ and $m_S$ + 1 with $-S \leq m_S \leq S - 1$ and $m_S \neq -1/2$. The broadening of the STs results from the first-order Zero-Field Splitting (ZFS) related to the electron-electron dipolar interactions, whereas the CT is only broadened by the second-order ZFS. In the case of Mn(II) ions, the CT consists of a sextet centered at g ≈ 2 owing to the hyperfine coupling with $^{55}$Mn nucleus with spin $I$ = 5/2. For these ions, the dominant DNP mechanism is the SE by irradiation of the CT. Nevertheless, a contribution of CE involving the electrons of STs has also been suggested [24,146,147].

Besides the above exogeneous PAs, which can be introduced into the investigated diamagnetic material, unpaired electrons of paramagnetic materials have also been used as endogeneous PAs. This approach has been used for (i) mesoporous silica functionalized with TEMPO, which is used as a selective oxidation heterogeneous catalysts [148], (ii) thermally carbonized mesoporous silica containing carbon dangling bonds [149], (iii) diamond microcrystals and nanoparticles with a diameter ranging from 45 to 80 nm [150] and (iv) silicon nanoparticles containing dangling bonds [133].

### 3.2. Sample preparation

The DNP-NMR of diamagnetic materials requires the introduction of PAs in the sample, without diluting the sample, in order to preserve the sensitivity. For instance, PAs have been introduced into porous materials using incipient wetness impregnation [36,51]. In that technique, the volume of radical solution is just sufficient to fill the pores (see Figure 7a). A similar incipient wetness wetting technique has been used for particulate materials. The volume of radical solution is just enough to uniformly wet the surface of the particles. For colloids, including sols, i.e. colloidal suspensions of solid particles in a liquid, and gels, which are solid three-dimensional networks expanded by a liquid, the PA can be dissolved into the liquid phase [85,151,152]. Furthermore, soluble polymers can be swollen in a radical solution [153].

For these techniques, the choice of the solvent can strongly affect the DNP enhancement. The employed solvent should fulfill the following criteria: (i) it should dissolve the PA; (ii) it should wet the investigated material, (iii) it should be inert towards both the radical and the material and must not dissolve it; (iv) the number of magnetically inequivalent nuclei in the solvent must be low in order to avoid masking the spectrum of the material; (v) the solvent must form a glass at the temperature of DNP experiments; (vi) the solvent molecules should not exhibit sizable motions, such as the rotation of methyl groups, at the desired temperatures; in addition, for the DNP involving protons, (vi) the concentration of protons in the solvent must be less than 20 mol.L$^{-1}$ [64,129] and (vii) the longitudinal relaxation of the solvent protons should be slow.

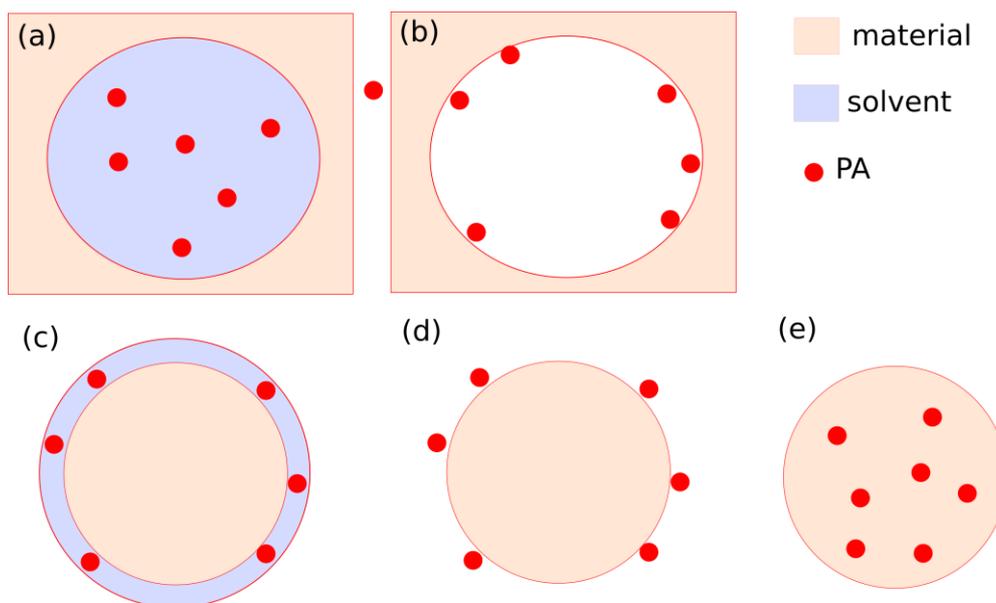

**Figure 7.** Schemes depicting the spatial distributions of material, PAs and solvent for DNP samples of (a, b) porous and (c-e) particulate materials, in which PAs are incorporated using (a, c) incipient wetness impregnation or wetting, (b, d) the adsorption of PAs on material surface and (e) the doping of material bulk with PAs.

The crystallization of the solvent must be avoided since it leads to aggregation of the PAs along the grain boundary, which decreases the DNP enhancement [47]. In practice, crystallization can be avoided by (i) the use of glass-forming solvents, such as water/glycerol and water/DMSO mixtures [41,154–156], (ii) the confinement of the solvent into micro- or meso-pores, which prevents the formation of large crystalline domains [36,157], and (iii) a rapid freezing of the solvent [137]. Similarly the aggregation and the precipitation of nanoparticles upon the freezing of sols can result in the segregation between aggregated nanoparticles and PAs [158]. The aggregation is governed by the steric hindrance of surface groups and their affinity for solvent molecules [118]. Such aggregation can be prevented by (i) the impregnation of mesoporous silica with a sol, so that PAs and nanoparticles are trapped in the same mesopore and cannot segregate [158,159] and (ii) the dispersion of the nanoparticles into a gel medium, instead of a liquid phase [152].

It has been shown that the partial deuteration of the solvent generally increases the $^1$H DNP enhancement [155] since (i) for partially deuterated solvents, the electron polarization is transferred to a reduced number of protons and a larger fraction of them are polarized and (ii) the reduced number of protons slows the electron and nuclear relaxations, which increases the efficiency of the DNP transfer. Nevertheless, the reduced $^1$H-$^1$H dipolar couplings lead to slower spin diffusion, which can produce polarization losses, notably at faster MAS [98]. The deuteration of the solvent also reduces the intensity of the solvent signal for DNP via $^1$H, when the solvent contains both hydrogen atoms and the detected isotope [153,160]. For the direct polarization of isotopes other than proton using nitroxide PAs, the total deuteration of the solvent can be advantageous in order to limit the transfer of polarization to the protons [140]. Nevertheless, optimal $B_0$ fields for DNP of $^2$H and other isotopes with low γ ratio are close and the polarization transfer to $^2$H can compete with that to the detected isotope. The optimal concentration of protons depends on the PA, the involved DNP mechanism [161], the MAS frequency [98] and the concentration of protons in the investigated sample [160]. Typically the concentration of protons in the solvent must be on the order of 10 mol.L$^{-1}$ [64]. For water-soluble PAs, such as AMUPol, [$^2$H$_8$]-glycerol/$^2$H$_2$O/H$_2$O (60/30/10 v/v/v) or [$^2$H$_6$]-DMSO/$^2$H$_2$O/H$_2$O (80/10/10 v/v/v) solvents are typically used. For radicals soluble in organic solvents, such as TEKPol, the most widely used solvent is isotopically unmodified TCE, which has a $^1$H concentration of 20 mol.L$^{-1}$ [64,129]. However, other solvents, such as [$^2$H$_2$]-TCE [140,153], C$^2$HCl$_3$/[$^2$H$_2$]-TCE/TCE (20/24/56 v%) [162], 1,1,2,2-tetrabromoethane [124,129], CHCl$_3$/1,1,1,2-tetrabromoethane/[$^2$H$_4$]-methanol (65/30/5 v%) [83] or [$^2$H$_{14}$]-OTP/OTP [69,94,95,163] have been used.

The impregnation or wetting of the material of interest with a radical solution can have some drawbacks. The radical solution can modify the structure and the dynamics of the sample. For instance, it has been shown that for some forms of theophylline, an asthma drug molecule, sample grinding and impregnation with radical solution induce polymorphic transitions and desolvation [164]. The addition of radical solution can also (i) dilute the investigated material, and hence, reduce the effective amount of sample in the rotor as well as (iii) broaden the NMR signals owing to the distribution of local environments in the frozen sample and at its surface. The DNP-enhanced signals of the solvent, notably for $^1$H and $^{13}$C, can also mask signals from the material of interest [165,166].

To circumvent the aforementioned issues, techniques have been developed to study materials using DNP-NMR with a minimized amount of the frozen solvent. The first minimal-matrix approach for MAS DNP was the polymer solution casting, in which the polymer was dissolved in radical solution and subsequently the solvent was evaporated [153,167,168]. It has been shown that this method produces narrower lines than the swelling in radical solution for the study of semi-crystalline polymers [153]. However, for polymer solution casting, the slow evaporation of the solvent may cause the exclusion of the radicals from the polymer crystalline domains, which results in lower sensitivity than for impregnation methods.

Another approach to minimize the amount of solvent involves mixing the material of interest with a radical solution, which does not solubilize the material, followed by the evaporation of the solvent [121,160]. The mixing with the radical solution can consist of the preparation of a suspension for nanoparticles [121] or the impregnation for porous materials [160]. This method is particularly efficient for materials on which the PA adsorbs because this adsorption prevents the aggregation of the PA during the evaporation of the solvent (see Figure 7b and d) [55]. In particular, it has been shown that some nitroxide biradicals can bind to sugars, such as cellulose or peptidoglycans, [121,169] as well as to silica or alumina surfaces via hydrogen bonding (see Figure 8c) [124,170].

Nitroxide radicals have also been incorporated into organic materials using minimal-matrix preparation methods, such as co-precipitation [171], spray drying and hot-melt extrusion [163,172]. In spray drying, a solution containing the investigated molecules and the radicals is sprayed against a flow of warm air, which rapidly evaporates the solvent and yields a powder. In hot-melt extrusion, the solid constituents are mechanically mixed at elevated temperatures. Paramagnetic metal ions have been incorporated into the inorganic solids using co-crystallization [71] or *via* solid-state reactions [72,73].

Another important parameter is the concentration of the PA, [PA]. For low [PA], the sensitivity increases with increasing [PA] since (i) each unpaired electron has to polarize a lower number of nuclei and (ii) the build-up of the DNP-enhanced nuclear polarization is faster. However, paramagnetic quenching and the shortening of coherence lifetimes, which produce losses during the pulse sequence, result in a decrease of sensitivity with higher [PA]. Furthermore, intermolecular electron-electron dipolar couplings tend to equalize the electron polarization through the EPR line [79,90]. Therefore, there is an optimal concentration, which depends on the nature of the PA, the investigated material and the employed pulse sequence [123,124,170]. Materials with greater surface areas require higher [PA], when the PAs adsorb on the surface of the material (see Figure 8) [170]. In addition, low [PA] is preferred for long polarization transfers [123]. In the case of material impregnation with nitroxide biradicals, the [PA] producing highest sensitivity can range from a few to 30 mmol.L$^{-1}$ [170].

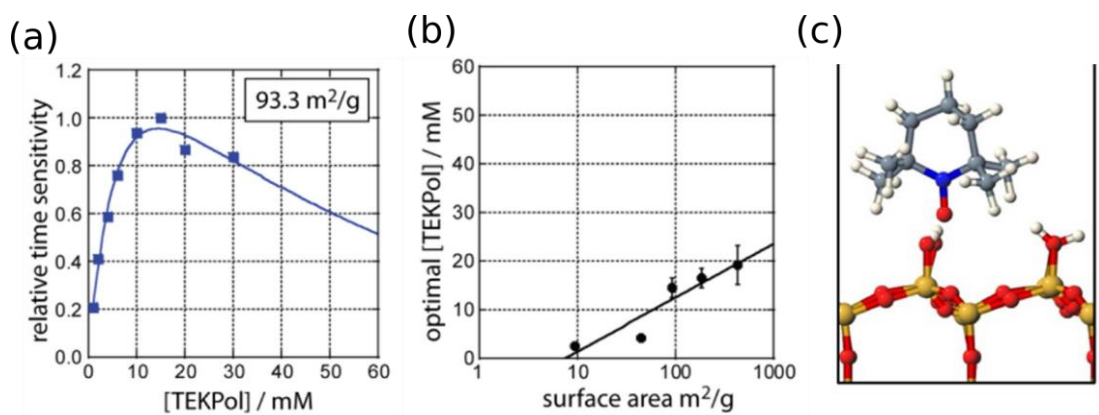

**Figure 8.** (a) Plot of the relative sensitivity of $^1$H→$^{29}$Si CPMAS experiments on non-porous silica nanoparticles with a BET surface area of 93.3 m$^2$.g$^{-1}$ wetted by a TEKPol solution in TCE as function of the TEKPol concentration. (b) Plot of the TEKPol concentration yielding the highest sensitivity as function of the surface area. (c) The most stable adsorption structure of a TEMPO molecule on β-SiO$_2$ (111) surface calculated using DFT. The H, C, N, O and Si atoms are represented using white, gray, blue, red and brown spheres, respectively. Figure adapted with permission from ref. [170]. Copyright 2018, Elsevier.

## 4. Instrumentation and experimental parameters

### 4.1. Instrumentation

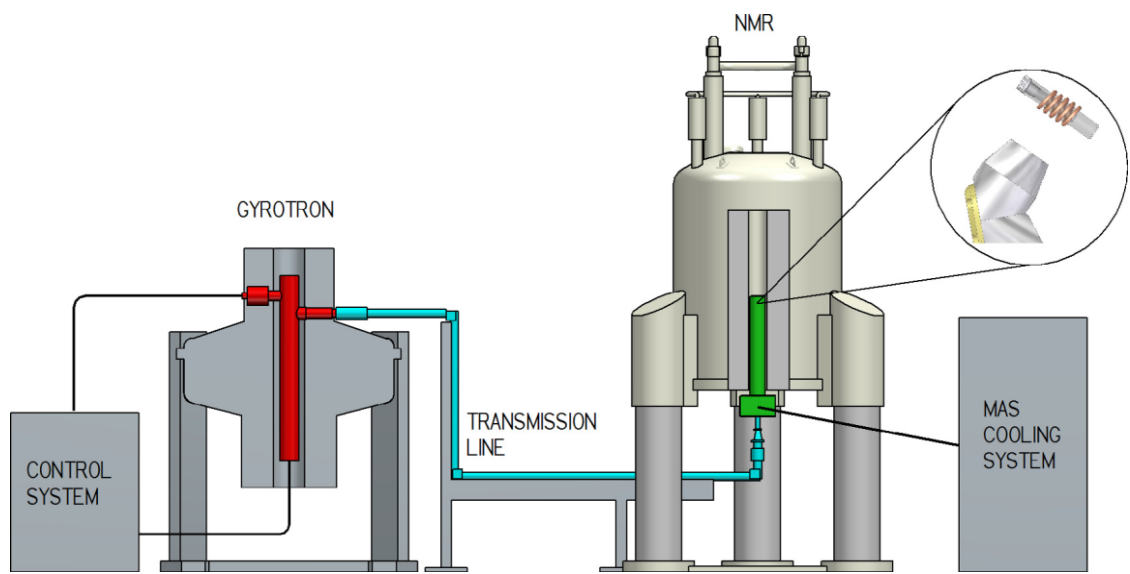

**Figure 9.** High-field MAS DNP-NMR system with gyrotron microwave source (gyrotron tube in red), microwave transmission line (cyan) and low-temperature NMR probe (green). Figure adapted with permission from ref. [56]. Copyright 2016, Elsevier.

Most high-field MAS DNP-NMR experiments on materials have been carried out on commercial systems manufactured by the Bruker BioSpin company (see Figure 9) [43,56]. Therefore, the MAS DNP-NMR instrumentation is only briefly discussed here. Further information on that topic can be found elsewhere [24,43,46,56,58]. The microwave source is a continuous-wave gyrotron with a dedicated magnet [39,43,58]. The microwave beam delivered by the gyrotron propagates through a transmission line, which is a corrugated waveguide [173]. This waveguide terminates close to the radiofrequency (rf) coil of the MAS DNP-NMR probe, where the sample is located. The microwaves are launched between the turns of the sample rf coil. Besides conventional solid-state NMR capabilities and microwave transmission, the MAS DNP-NMR probe must be suitable for MAS at low temperature. This functionality is accomplished by blowing on the rotor not only cold bearing and drive gas, but also an additional cooling gas. A heat exchanger is employed to cool the gas [174,175], which is transported to the MAS stator by thermally insulated transfer lines. Furthermore, the probe can be equipped with a pneumatic insert and eject system, which allows the change of the sample at low temperature [176]. The probe is placed within an NMR magnet. Since commercially available gyrotrons operate at fixed microwave frequency, the NMR magnet can be equipped with an additional superconducting sweep coil, in order to tune the static magnetic field at the position of the sample and match the condition for maximal DNP enhancement. Note that the development of frequency tunable gyrotrons would alleviate the need to sweep the field of the NMR magnet [177]. However, the output power of these microwave sources varies significantly as function of the frequency and their use for DNP-NMR experiments has been limited so far.

### 4.2. Microwave source and microwave power

As explained in section 2.1, an increase of the microwave power improves the adiabacity of microwave crossings and hence, the DNP efficiency. Simulations and experiments show that the DNP signal enhancement increases with increasing microwave field strength up to a plateau [43,56,74,76,178]. This plateau corresponds to the saturation of the allowed EPR transition for the OE and that of forbidden zero-quantum or double-quantum transitions for the SE, *i.e.*, equal populations of the two energy levels of the transition. In the case of the CE, this saturation corresponds to the case, where the polarization of one electron is close to zero, whereas the other is at thermal polarization [74]. The required microwave field strength to reach saturation depends on the DNP mechanism as well as on the longitudinal and transverse electron relaxation times of the PA [24,74,179]. In the case of the CE using TOTAPOL biradicals at 9.4 T and 100 K, simulations predict a required microwave field strength on the order of 1 MHz [74,76].

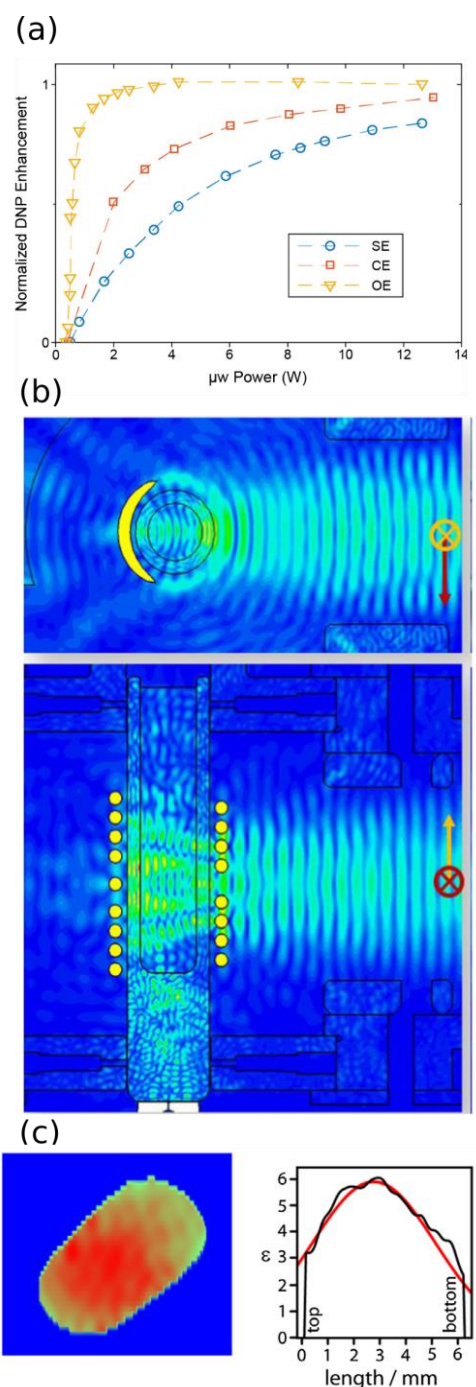

**Figure 10.** (a) Normalized DNP enhancement as function of microwave power for the OE, the SE and the CE DNP. (b) Simulated spatial distribution of the microwave magnetic field in a radial plane (top) and an axial plane (bottom) of 3.2 mm sapphire rotor in 263 GHz MAS DNP-NMR system. The microwave electric field is polarized parallel to the rotor axis and is depicted as a yellow cross in the radial plane and a yellow arrow in the axial plane. (c) $^1H \rightarrow {}^{31}P$ CP-STRAFI-MAS 2D image of DNP enhancement (left) and cross-section of the image along the rotor axis (right) for $Na_3PO_4 \cdot 12H_2O$ wetted with a 16 mM TEKPol solution in TCE at 9.4 T and 100 K with $\nu_R$ = 10 kHz. Figure adapted from refs. [24,56,180]. Copyright 2016 and 2017, Elsevier.

For high-field MAS DNP experiments, the size of sample, which is in the order of a few millimeters, is comparable to the wavelength of the microwaves, which ranges from 1.1 to 0.5 mm for electron Larmor frequencies ranging from 263 to 593 GHz. These wavelengths preclude the use of DNP-NMR probes with microwave resonant structure. Therefore, high-power microwave sources are required for MAS DNP-NMR experiments. In the frequency range corresponding to high-field MAS DNP (263–593 GHz), gyrotrons are still the sources producing the highest microwave power [46,58]. They can deliver high-power microwaves at fixed frequency with power stability better than ±1% for several days, i.e. the time required for long solid-state NMR experiments. A drawback of gyrotrons is the need for a superconducting magnet. Nevertheless, the use of microwaves at the second harmonic of the electron cyclotron frequency, instead of the first harmonic, allows the strength of the static magnetic field for the gyrotron magnet to be divided by a factor of two [181]. Second harmonic gyrotrons using cryogen-free superconducting magnets of 5, 7, 10 and 11.1 T have been developed by Bruker and CPI to produce continuous microwave at frequencies 263, 395, 527 and 593 GHz, respectively, for DNP-NMR experiments at 9.4, 14.1, 18.8 and 21.1 T, respectively (corresponding to $^1$H Larmor frequencies of 400, 600, 800 and 900 MHz) [56,58,182]. These gyrotrons can typically deliver microwaves with a maximal output power of 50 W.

At the end of the waveguide, the microwaves are diffracted and reflected by the rf coil, which has a pitch comparable to the wavelength of the microwaves. Because of the reflection of the microwaves, standing waves occur and display patterns with a period related to the wavelength [56,178,183]. Hence, the strength of the microwave magnetic field is highly inhomogeneous within the sample space. This inhomogeneous distribution of the microwave field results in an inhomogeneous distribution of the DNP enhancement, the enhancement factors being larger at the center of the rotor, where the microwave beam irradiates the sample (see Figure 10c) [180]. This result was obtained by mapping the DNP enhancement in the sample using stray-field imaging under MAS (STRAFI-MAS) experiments.

Furthermore, this strength depends on the efficiency of the conversion of the incident microwave power into microwave field amplitude. This conversion depends on several factors, such as the design of the DNP-NMR probe (coating of the inner surface of the stator, pitch of the sample rf coil, use of a lens to focus the microwave beam on the rotor [139,178], etc.), the rotor and the investigated sample. For instance, sapphire ($Al_2O_3$) rotors are more transparent to microwaves than zirconia ones. For 263 GHz DNP-NMR system equipped with 3.2 mm sapphire rotors, it has been shown that the average electron nutation frequency, which is related to the amplitude of the microwave field, is about 0.2–0.3 MHz [83,87]. Sapphire rotors also benefit from a higher thermal conductivity, which limits the heating of the sample during microwave irradiation [43]. Nevertheless, thin-wall zirconia rotors can be utilized to avoid observing $^{27}$Al background signal of sapphire rotors [184,185]. The penetration of microwaves and hence, the DNP enhancements depend on the thickness of the rotor walls and the sample [47,186]. Furthermore, it has been shown that the addition of particles, such as KBr or sapphire, with high real and low imaginary parts of the permittivity, gives rise to additional diffraction and reflection of the microwaves in the sample space and so amplifies the amplitude of the microwave field in the sample, which increases the DNP enhancement [87]. It has been shown that this technique can improve the sensitivity of DNP experiments for static or slowly rotating samples [112,180]. However, the addition of these particles significantly reduces the active sample volume and generally does not provide a global sensitivity gain under MAS, at least when the available volume of sample is sufficient to fully fill the rotor.

Conversely, low DNP enhancements can be obtained when the investigated material has unfavorable dielectric properties. In particular, materials exhibiting a high imaginary part of the permittivity absorb the electric field energy of the microwave and dissipate it as heating, which increases the temperature of the sample, accelerates the electron and nuclear relaxations and hence, reduces the efficiency of the DNP transfer [123,187]. This microwave-induced heating is especially large for conductive samples. It also increases with increasing microwave power, which limits the maximal microwave power, which can be employed in practice [178]. Furthermore, the imaginary part of the permittivity usually increases at higher frequencies and hence, the heating due to microwave is increased for DNP at higher $B_0$ field.

### 4.3. Magnetic field of the NMR magnet

Higher static magnetic fields improve the resolution of NMR spectra, when the linewidth is not dominated by the distribution of isotropic chemical shifts produced by the disorder in the material or the anisotropy of bulk magnetic susceptibility [188]. For spin-1/2 nuclei exhibiting field-independent linewidths, the resolution is proportional to $B_0$. For half-integer spin quadrupolar nuclei, the CT linewidth is dominated by the second-order quadrupolar broadening, which is inversely proportional to $B_0$, and the resolution increases as $(B_0)^2$.

Furthermore, in the case of conventional solid-state NMR experiments, high $B_0$ field generally improves the sensitivity since it enhances the thermal polarization and the induced voltage in the sample coil and also reduces the CT second-order quadrupolar broadening for half-integer quadrupolar nuclei. Hence, even if the electronic noise increases with higher field owing to the skin effect, the signal-to-noise ratio increases as $(B_0)^{7/4}$ for spin-1/2 nuclei and $(B_0)^{11/4}$ for half-integer quadrupolar nuclei [189]. The dependence of the sensitivity on $B_0$ may depart from these relationships, when the longitudinal nuclear relaxation time depends on $B_0$ or a significant chemical shift anisotropy (CSA) reduces the sensitivity gain at higher $B_0$ field by enhancing spinning sidebands and lowering the centerband intensity.

In addition, in the case of DNP-NMR experiments, the efficiency of DNP transfer usually depends on the $B_0$ field. As explained in section 2.1, the efficiency of the OE in solids has been shown to increase with higher $B_0$ field, whereas those of the SE and the CE decrease with increasing $B_0$ field, the decrease being faster for the SE [76]. For the SE and the CE, the decrease of the efficiency at higher field stems from the broadening of the EPR line, which is approximately proportional to $B_0$. This broadening decreases the adiabacity of the level crossings. Despite its unfavorable dependence on $B_0$, the CE mechanism still yields the highest sensitivity at $B_0$ = 18.8 T [69]. Furthermore, for spin-1/2 nuclei in frozen solutions, it has been shown that the CE using TEMTriPol-1 nitroxide-trityl bidradicals can yield a sensitivity at 18.8 T comparable to that obtained 9.4 T [138]. Therefore, MAS DNP-NMR experiments on half-integer spin quadrupolar nuclei using TEMTriPol-1 as PA should be more sensitive at 18.8 T than at 9.4 T. Hence, the development of PAs exhibiting narrow EPR lines is desirable for MAS DNP NMR at $B_0 > 10$ T [139].

Currently, the majority of MAS DNP-NMR experiments on materials has been carried out at $B_0$ = 9.4 T. Nevertheless, some MAS DNP-NMR experiments on materials have been reported at 14.1 T [158,159,190–197], 18.8 T [69,80,94,95,139,186] and 21.1 T [139,198]. DNP-NMR experiments at $B_0 \geq 14.1$ T benefit notably from a higher spectral resolution for the detection of half-integer quadrupolar nuclei. For instance, the $^{27}$Al signals of tetra- ($AlO_4$), penta- ($AlO_5$) and hexa-coordinated ($AlO_6$) surface sites of γ-alumina are better resolved at 18.8 T [80,139] than at 9.4 T [124,184].

### 4.4. Temperature

As explained in section 2, slow electron and nuclear relaxations improve the efficiency of the DNP transfer. Furthermore, slow nuclear relaxation facilitates the transport of DNP-enhanced magnetization within the sample by nuclear spin diffusion and reduces the losses during the pulse sequence. Therefore, higher DNP enhancements and higher sensitivity are achieved at lower temperature [43,69,123,130,137,179]. Microwave irradiation and MAS can heat the sample. Currently, almost all MAS DNP-NMR experiments on materials have been carried out at about 100 K using cold $N_2$ gas to cool the sample and spin it. MAS DNP-NMR experiments at lower temperatures, down to 6 K, have also been reported using helium instead of nitrogen gas and it has been shown that they are more sensitive than those at *ca*. 100 K [19,21,62,179,199–202].

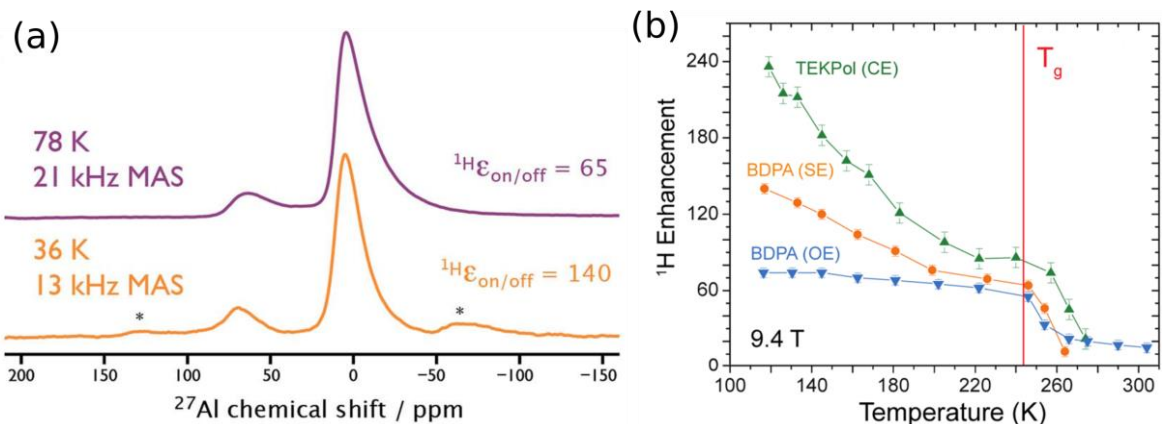

**Figure 11.** (a) $^1H{\rightarrow}^{27}Al$ CPMAS spectra of γ-alumina impregnated with 10 mM AMUPOL solution in $[^2H_6]$-DMSO/$^2H_2O$/$H_2O$ (78/14/8 w/w/w) at 9.4 T and 78 (top, purple) and 36 K (bottom, orange) with MAS frequencies of 21 and 13 kHz, respectively. Asterisks denote spinning sidebands. (b) $^1H$ DNP enhancement as function of temperature for 32 mM BDPA (enhancement maxima corresponding to SE and CE) or 16 mM TEKPOL solution in $[^2H_{14}]$-OTP/OTP (95/5 mol/mol) mixture at 9.4 T with $\nu_R$ = 8 kHz. Figure adapted from refs. [69,179]. Copyright 2015, Royal Society of Chemistry and American Chemical Society.

Nevertheless, the need for low temperature can be a limitation for the characterization of materials, notably to study their dynamics. Even if the sensitivity enhancement produced by DNP decreases with increasing temperature, it has been shown that DNP can still yield significant sensitivity enhancements at temperatures much higher than 100 K [69,137,150]. For instance, a DNP enhancement of *ca*. 15 has been reported at room temperature using the OE of BDPA in OTP solvent [69]. Furthermore, DNP has been shown to yield at room temperature enhancements as large as 140 for diamond microparticles containing endogeneous paramagnetic centers under MAS [150] and 1500 for synthetic diamond single crystals under static conditions [203]. Rigid samples with limited molecular motions, and hence, slower electron and nuclear relaxation exhibit higher DNP enhancements at high temperatures.

*4.5. MAS frequency*

MAS improves the resolution of solid-state NMR spectra by averaging out the anisotropic interactions. For instance, the suppression of spinning sidebands due to CSA or second-order quadrupolar interaction requires MAS frequencies larger than these interactions. Fast MAS also improves the resolution of $^1H$ spectra since the width of individual $^1H$ lines is generally inversely proportional to the MAS frequency [188].

On the other hand, MAS can also affect the efficiency of the DNP transfer and the nuclear spin diffusion. This dependence has so far mainly been investigated for the CE [74–76]. Under static conditions, the CE can only occur if the two unpaired electrons and the nucleus match the CE condition $|\nu_0(S_a) - \nu_0(S_b)| \approx |\nu_0(I)|$ and the microwave irradiation is resonant with one of the electrons. Hence, only a limited number of PA molecules contributes then to the CE transfer during the whole microwave irradiation. MAS sweeps the resonance frequencies of the unpaired electrons and allows a larger number of PA molecules to contribute then to the CE transfer during the level anti-crossings. Furthermore, as the microwave field is highly inhomogeneous (see Figure 10b), MAS ensures a more uniform microwave irradiation of the sample [43,56]. Additionally, the MAS CE mechanism described in subsection 2.1 is only efficient when the multiple energy level crossings and anti-crossings occur within a time shorter than $T_1(S_i)$ [74]. As an example for nitroxide biradicals, such as AMUPol, $T_1(S_i)$ times are about 0.5 ms [83] and as a result, the efficiency of the DNP transfer increases up to a MAS frequency of *ca.* 2 kHz (see Figure 12a). As more PAs take part in the CE transfer under MAS conditions than under static ones, the build-up of the DNP-enhanced polarization is faster under MAS [80].

As explained in subsection 2.1, larger MAS frequencies also decrease the adiabaticity of the crossings and anti-crossings and hence can reduce the efficiency of the DNP transfer, as seen in Figure 12a for AMUPol. It has been shown that such a deleterious effect can be avoided by using PA featuring significant exchange and/or dipolar interactions, such TEMTriPol-1, AsymPolPOK or HyTEK2 [80,138,139].

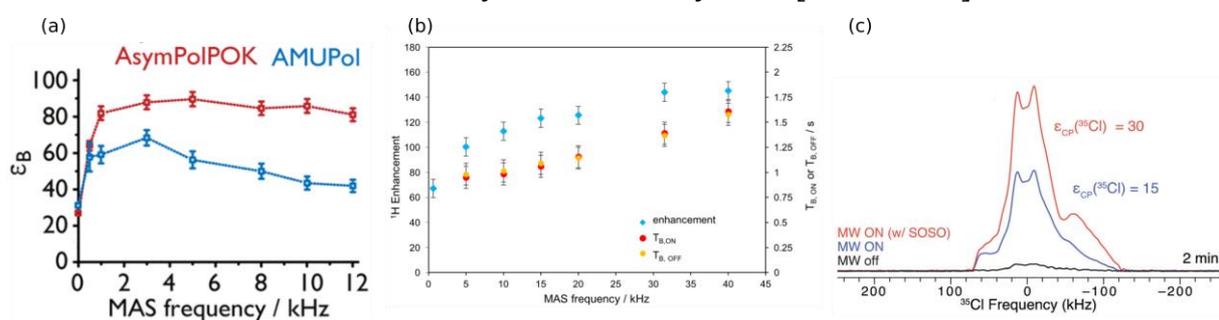

**Figure 12.** (a) Enhancement of the $^1$H polarization compared to thermal one as function of the MAS frequency for 10 mM AsymPolPOK (red) and AMUPol (blue) solutions in [$^2$H$_8$]-glycerol/$^2$H$_2$O/H$_2$O (8/1/1 v/v/v) mixture at 9.4 T and 105 K. (b) Enhancement of $^1$H polarization and build-up time constants of $^1$H polarization with ($T_{B,ON}$) and without microwave irradiation ($T_{B,OFF}$) as function of MAS frequency for 32 mM HyTEK-2 solution in TCE at 9.4 T and 125 K. (c) $^1$H→$^{35}$Cl broadband adiabatic inversion cross-polarization (BRAIN-CP) spectra with Carr-Purcell Meiboom-Gill (CPMG) detection using wideband uniform-rate smooth truncation (WURST) pulses of finely ground ambroxol hydrochloride, a drug used in the treatment of respiratory diseases, impregnated with 15 mM TEKPol solution in TCE at 9.4 T and 100 K. These spectra were acquired with microwaves and MAS rotation during most of the polarization delay but static conditions during the BRAIN-CP-WURST-CPMG sequence (SOSO acquisition, red), with microwaves and stationary sample (blue), and without microwaves and with stationary sample (black). Figure adapted from refs. [80,139,204]. Copyright 2016, Royal Society of Chemistry, 2018, American Chemical Society.

Furthermore, for several radicals, such as BDPA or HyTEK2, it has been observed that the build-up of the DNP-enhanced polarization is slower for increasing MAS frequency (see Figure 12b) [94,139]. This slower build-up has been ascribed to slower $^1$H-$^1$H spin diffusion. It has also been shown that for $^{119}$Sn nuclei subject to large CSA, the sensitivity can be maximized by using low MAS frequencies during the polarization delay in order to accelerate the spin diffusion and fast MAS frequencies during the acquisition to decrease the intensity of the CSA spinning sidebands [205].

From a practical point of view, the higher density of N$_2$ gas and the reduced speed of sound in N$_2$ for the temperatures close to its liquefaction point (77 K at atmospheric pressure) result in significantly reduced maximum MAS frequencies compared to MAS experiments at room temperature. So far the majority of MAS DNP NMR experiments have been carried out using 3.2 mm rotors cooled and spun by cold nitrogen gas. In such cases, at temperatures of about 100 K, the maximum MAS frequency of these rotors cannot exceed 15 kHz [43]. It has been shown that a maximum MAS frequency of 25 kHz can be reached at 90 K by cooling and spinning these 3.2 mm rotors using cold He gas (see Figure 11a) [179]. Furthermore, MAS DNP NMR probes for 1.3 mm rotors have recently become available and have enabled DNP NMR experiments at MAS frequencies up to 40 kHz at 100 K and 9.4 and 18.8 T [94,139,186].

Although this review focuses on MAS DNP-NMR of materials, it is appropriate to mention that high-field DNP NMR experiments on materials under static conditions have been recently reported [63,112,204,206,207]. These experiments are notably useful to probe the orientation of molecules in aligned samples [206] or when the magnitude of the anisotropic interactions, such as CSA or second-order quadrupolar interactions, greatly exceed the accessible MAS frequencies [112,204,207]. These experiments have been carried out using specific static DNP-NMR probes with a sample coil orthogonal to the $B_0$ field [206] or MAS DNP-NMR probe [112,204,207]. However, as explained above, CE DNP yields lower sensitivity gains when it occurs under static condition, instead of MAS. This issue can be circumvented by applying MAS during the polarization delay and stopping it during the NMR pulse sequence. This approach, called spinning-on spinning-off (SOSO) acquisition, has been shown to yield a two-fold higher signal enhancement over the acquisition under purely static conditions (see Figure 12c) [204].

## 5. Isotopes and NMR methods

### 5.1. Isotopes and type of transfer

**Figure 13.** Periodic tables showing the spin-1/2 (blue) and quadrupolar (red) isotopes, which have been detected using high-field MAS DNP-NMR experiments using either (a) coherent transfer of DNP-enhanced $^1$H polarization to the detected isotope or (b) direct excitation experiments.

Figure 13 shows the chemical elements that have been detected in materials using high-field MAS DNP-NMR. The sensitivity enhancement provided by DNP has notably been used to detect insensitive spin-1/2 isotopes, owing to (i) their low natural abundance (*NA*), such as $^{29}$Si (*NA* = 4.7%) [45,208], $^{13}$C (*NA* = 1.1%) (see Figure 1c) [36,209] or $^{15}$N (NA = 0.37%) [111], (ii) their low gyromagnetic ratio, such as $^{15}$N ($\gamma$($^{15}$N) ≈ 0.1$\gamma$($^{1}$H)) or $^{89}$Y ($\gamma$($^{89}$Y) ≈ 0.05$\gamma$($^{1}$H)) [109] or (iii) their large CSA, such as $^{77}$Se, $^{111}$Cd, $^{113}$Cd, $^{119}$Sn, $^{195}$Pt or $^{207}$Pb [112,119,151,158,207]. However, DNP-NMR has also been employed for receptive spin-1/2 nuclei, such as $^{31}$P [210].

MAS DNP-NMR has also been applied to detect spin-1 quadrupolar nuclei. For instance, the high-sensitivity of DNP has been used to detect $^{2}$H in natural abundance (*NA* = 0.01%) in organic solids, which gives access to information about hydrogen chemical shifts and $^{2}$H quadrupolar coupling [211]. MAS DNP has also been employed to enhance the weak signal of the forbidden overtone transition between the energy levels $m$ = +1 and −1 of the $^{14}$N nucleus [212]. MAS DNP NMR of $^{6}$Li isotope with *NA* = 7.59% has also been reported [73].

It has also been shown for materials that MAS DNP can enhance the NMR signals of insensitive half-integer spin quadrupolar nuclei with a low natural abundance, such as $^{17}$O ($I$ = 5/2, *NA* = 0.037%, |$\gamma$($^{17}$O)| ≈ 0.14$\gamma$($^{1}$H)) [105], and a low gyromagnetic ratio, such as $^{43}$Ca ($I$ = 7/2, *NA* = 0.14%, |$\gamma$($^{43}$Ca)| ≈ 0.07$\gamma$($^{1}$H)) [107] or with a low gyromagnetic ratio and subject to large quadrupolar interaction, such as $^{35}$Cl ($I$ = 3/2) (see Figure 11c) [204] and $^{79}$Br ($I$ = 3/2) [63]. MAS DNP has also been used to enhance the signals of receptive half-integer quadrupolar nuclei, such as $^{7}$Li ($I$ = 3/2) [72], $^{27}$Al ($I$ = 5/2) (see Figure 11a) [184], $^{51}$V ($I$ = 7/2) [213,214] and $^{59}$Co ($I$ = 7/2) [71].

Most of the MAS DNP experiments on materials employ a coherent transfer of the DNP-enhanced $^{1}$H polarization to the detected isotope. The phase cycle suppresses the direct DNP transfer from unpaired electrons to the detected isotope. This approach, termed coherent indirect DNP, has the advantage that the $^{1}$H-$^{1}$H spin diffusion transports the DNP-enhanced magnetization. Such transport acts to homogenize the $^{1}$H polarization within the sample and to accelerate its build-up. In the case of heterogeneous samples, in which protons and detected isotopes are located in different domains, the coherent transfer of $^{1}$H polarization to the detected isotope allows a selective observation of the nuclei near the interfaces since heteronuclear coherent transfers are only effective up to a few angstroms and the efficiency of transfer to remote nuclei is reduced by the dipolar truncation [215,216]. This strategy for the selective observation of surfaces and interfaces has been first demonstrated in the case of conventional NMR experiments [215]. It has later been shown that the sensitivity of these experiments can be enhanced using DNP at $B_0$ ≤ 1.5 T [35,38,167] and more recently at higher $B_0$ fields [36]. These surface-selective approach using indirect DNP has been dubbed surface-enhanced NMR spectroscopy (SENS) [36].For instance, this strategy has been used to selectively probe the surface of inorganic materials, for which the bulk region contains no or a limited amount of protons. The depth of coherent transfer below the surface depends on several parameters, including the concentration of the detected isotope, which notably depends on its natural abundance, its gyromagnetic ratio, the coherence lifetimes during the transfer and the duration of the coherence transfer [124]. For instance, it has been shown that DNP-enhanced $^{1}$H→$^{27}$Al CPMAS can be used to selectively observe the first layer of γ-alumina nanoparticles.

It has also been shown that $^{1}$H-$^{13}$C dipolar cross-relaxation can also transfer the DNP-enhanced $^{1}$H polarization to nearby $^{13}$C nuclei and produce a negative DNP enhancement, i.e. a DNP-enhanced $^{13}$C polarization pointing in opposite direction to the polarization at thermal equilibrium [146,217]. Such incoherent indirect DNP transfer was first reported for biomolecules [146,217] and more recently for nonionic surfactants, such as polyethylene glycol (PEG) and solutions of organic molecules, such as cyclohexane, in PEG (see Figure 14a) [209,218,219]. Local molecular motions, such as the rotation of the methyl groups or cyclohexane ring inversion, promote the $^1$H-$^{13}$C dipolar cross-relaxation. The rapid cooling of the samples and the formation of glass help to preserve the local molecular motions. This transfer can be suppressed by saturating the $^1$H transitions, for instance by applying a burst of π pulses on $^1$H channel, during the polarization build-up [209].

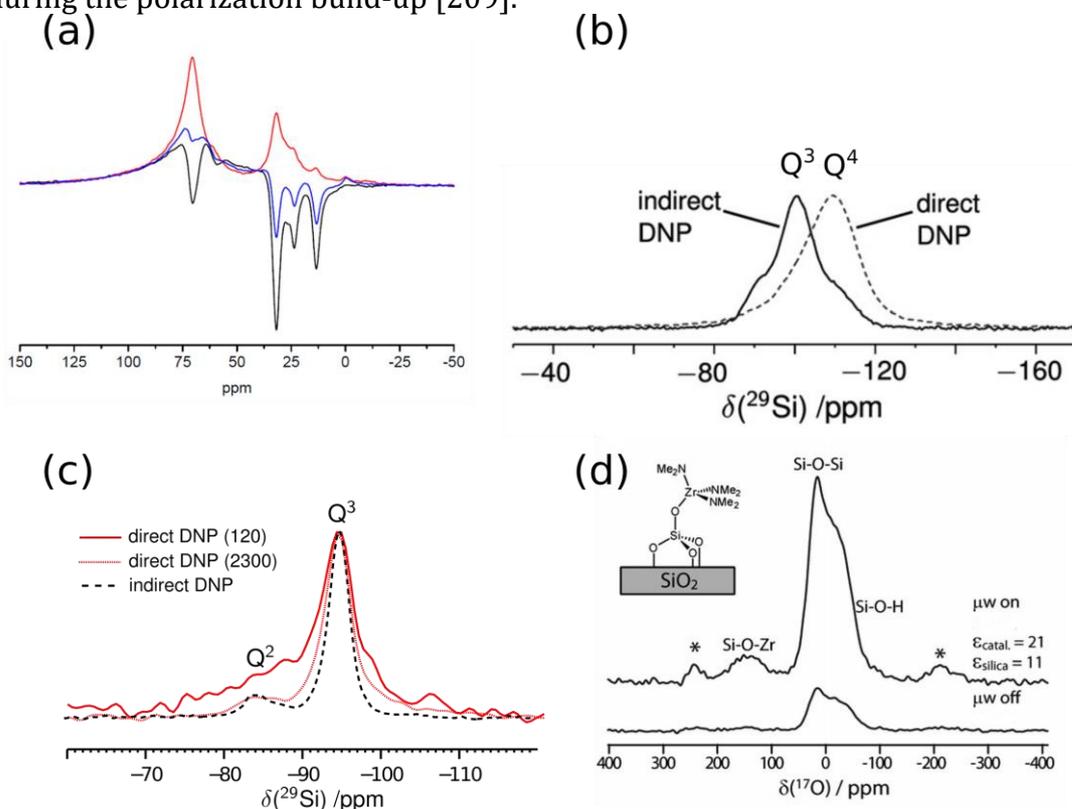

**Figure 14.** (a) $^{13}$C spectra of 20 mmol.kg$^{-1}$ AMUPol solution in nonionic surfactant with chemical formula HO[CH$_2$CH$_2$O]$_6$C$_{10}$H$_{21}$ with microwave irradiation at 9.4 T and 107 K with ν$_R$ = 8 kHz. The black spectrum was recorded using a $^{13}$C single-pulse sequence and is the sum of the direct DNP transfer and the incoherent indirect one. The red spectrum is only enhanced by direct DNP since the incoherent indirect transfer is suppressed by the application of π pulses on the $^1$H channel during the polarization delay. For the blue spectrum, 2π pulses, instead of π pulses, are applied on the $^1$H channel, which reintroduces the incoherent indirect transfer. (b,c) Comparison of the $^1$H→$^{29}$Si CPMAS and single-pulse $^{29}$Si spectra of (b) mesoporous silica impregnated with 15 mM TOTAPOL solution in [$^2$H$_6$]-DMSO/$^2$H$_2$O/H$_2$O (78/14/8 w/w/w) and (c) laponite synthetic clay dispersed in 20 mM TOTAPOL solution in the same mixture with microwave irradiation at 9.4 T and 100 K with ν$_R$ = 10 kHz. In both subfigures, the spectra are scaled to the same maximum intensity. In subfigure c, the single-pulse experiments were acquired with two distinct polarization delays, 120 s [direct DNP (120)] and 2300 s [direct DNP (2300)]. In the CPMAS and single-pulse spectra, the $^{29}$Si signals are enhanced by direct DNP and coherent indirect DNP, respectively. In subfigure b, the CPMAS and single pulse spectra are dominated by the signals of surface (SiO)$_3$SiOH ($Q^3$) sites and subsurface (SiO)$_4$Si ($Q^4$) sites, respectively. The subfigure c shows that direct DNP leads to increased linewidth with respect to indirect DNP but the line broadening of directly enhanced signals decreases with increasing polarization delay. (d) $^{17}$O NMR spectra of single-site catalysts, made of Zr(NCH$_3$)$_n$ species supported on mesoporous silica nanoparticles impregnated with 15 mM TEKPol solution in [$^2$H$_2$]-TCE with microwave irradiation (top) and without microwave irradiation (bottom) at 9.4 T and 100 K with ν$_R$ = 12.5 kHz. The signal of the $^{17}$O CT was enhanced by the application of a hyperbolic secant (HS) pulse before the CT-selective π/2 pulse. Figure adapted from refs. [45,85,140,209]. Copyright 2011, Wiley, 2013 and 2018, Royal Society of Chemistry, 2017, American Chemical Society.

An alternative to the indirect DNP consists of the direct transfer of polarization from unpaired electrons to the detected heteronucleus, i.e. an isotope distinct from proton. This approach has been termed direct DNP. In the case of materials, it has been applied for the detection of spin-1/2 nuclei, such as $^{13}$C [209], $^{15}$N [146], $^{29}$Si [45], $^{31}$P, $^{113}$Cd and $^{119}$Sn [119], but also of quadrupolar nuclei, such as $^6$Li [73], $^7$Li [72], $^{17}$O [105], $^{27}$Al [124], $^{51}$V [214] and $^{59}$Co [71]. Direct DNP complements the indirect version since it allows for the observation of heteronuclei more distant from protons. For instance, direct DNP has been used to observe subsurface or bulk sites in materials, for which protons are absent from the bulk (see Figure 14b and Figure 25) [45,72,73,119,195] as well as non-protonated surface $^{17}$O sites (see Figure 14d) [140]. Furthermore, as the dipolar couplings between heteronuclei are smaller than among protons, spin diffusion among heteronuclei is much slower than among protons. Hence, direct DNP often mainly polarizes the nuclei close to unpaired electrons. This heterogeneous polarization increases the linewidths with respect to conventional NMR spectra or those enhanced by indirect DNP (see Figure 14c) [85]. Furthermore, the build-up of the DNP-enhanced polarization cannot be fitted to a single exponential function but rather to a stretched exponential one (see Eq. 2) since the polarization build-up is slower for heteronuclei at increasing distance from the unpaired electrons [85,195]. However, in the case of slowly relaxing nuclei, it has been shown that nuclear spin diffusion can contribute to the transport of the DNP-enhanced polarization [119]. The transfer depth of direct DNP increases with increasing polarization delay and can reach distances of up to 10 nm for $^{119}$Sn [85,119,195,220].

## 5.2. NMR methods

Most MAS NMR pulse sequences can be used in combination with DNP. Nevertheless, the efficiency of the sequences can be reduced by the presence of PAs in the sample, which accelerates the relaxation and increases signal losses [123]. For experiments using coherent indirect DNP via protons, the excited nucleus must be proton and the DNP-enhanced $^1$H polarization has to be transferred to the other isotopes of interest. Such transfer is typically performed using the CPMAS sequence. It has been shown that the sensitivity can be further improved by the use of (i) the flip-back recovery method to utilize the residual $^1$H magnetization at the end of CW $^1$H heteronuclear decoupling [198], (ii) multiple contact CP [119,205,221] or (iii) the CPMG detection scheme for NMR signals inhomogeneously broadened by the distribution of isotropic chemical shifts, CSA or second-order quadrupolar broadening [120]. However, the CPMAS technique can lack robustness and efficiency for half-integer quadrupolar nuclei. For the transfer DNP-enhanced $^1$H polarization to surface $^{17}$O or $^{27}$Al sites, it has been shown that this issue can be circumvented by using PRESTO (Phase-shifted Recoupling Effects a Smooth Transfer of Order) technique, for which no spin lock is applied to the quadrupolar nucleus (see Figure 15) [222,223]. For isotopes exhibiting broad NMR signals, the coherent transfer of polarization can be achieved under static conditions using the BRAIN-CP technique (see Figure 12c) [112].

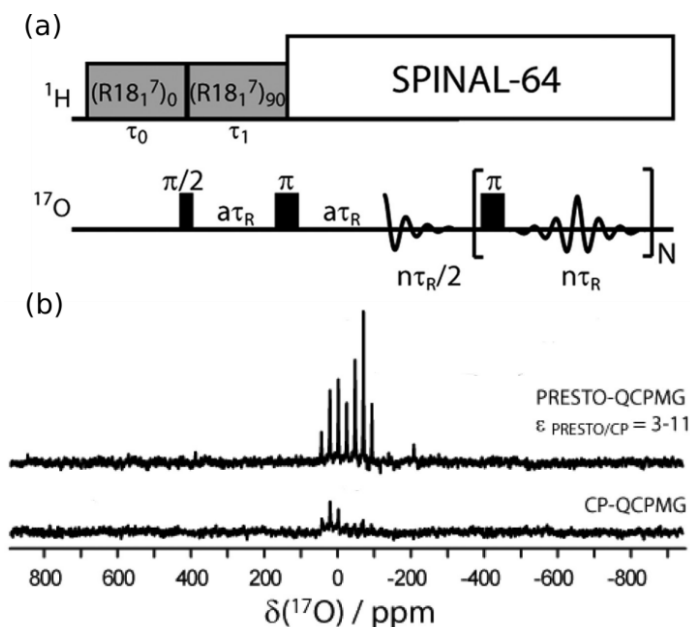

**Figure 15.** (a) $^1H \rightarrow ^{17}O$ PRESTO-QCPMG pulse sequence, where $\tau_R$ denotes the rotor period. (b) Comparison of $^1H \rightarrow ^{17}O$ PRESTO-QCPMG and CP-QCPMG spectra of $Ca(OH)_2$ wetted with 16 mM TEKPol solution in TCE with microwave irradiation at 9.4 T and 105 K with $\nu_R$ = 12.5 kHz. The PRESTO scheme yields a 3 to 11-fold enhancement in sensitivity with respect to CP. Figure adapted from ref. [222]. Copyright 2015, American Chemical Society.

As mentioned in section 3.2, a solvent is often added to the investigated material for DNP-NMR experiments. Even when the solvent is partially deuterated, its $^1H$ and $^{13}C$ signals can mask those of the investigated materials. These solvent signals can be suppressed by various methods, including (i) the use of short contact times in CPMAS to selectively transfer the DNP-enhanced $^1H$ polarization to protonated $^{13}C$ nuclei of the investigated materials and not to deuterated $^{13}C$ sites of the solvent, (ii) the reintroduction of $^{13}C$-$^2H$ dipolar couplings, (iii) the use of relaxation filters, since for heterogeneous samples, such as a solid particles wetted by a PA solution, the nuclei in the solvent relax faster than those of the investigated material owing to their closer proximity to PA, and (iv) the impregnation of the sample with highly concentrated radical solutions, which results in rapid spin-lattice relaxation in the rotating frame and hence, low efficiency of the CPMAS transfer for the molecules of the frozen solvent [165,166,224].

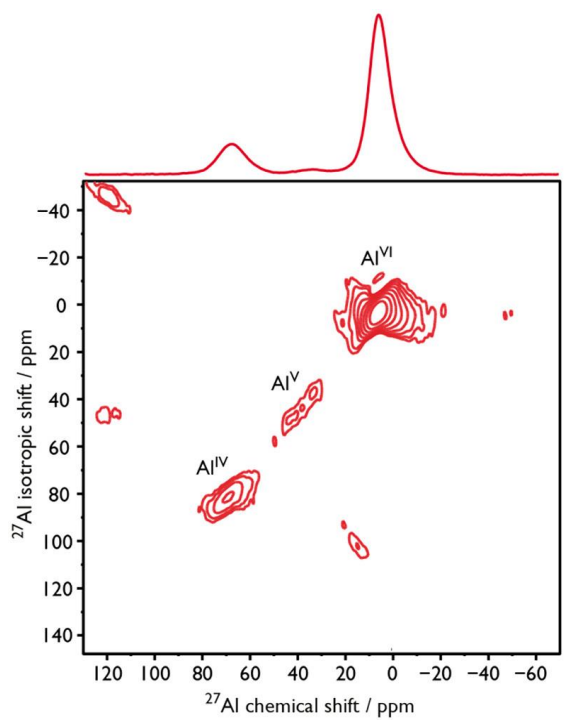

**Figure 16.** $^1H \rightarrow ^{27}Al$ CP-3QMAS 2D spectrum of γ-alumina impregnated with a solution of 5 mM AsymPolPOK and 2 M [$^{13}C$]-urea in [$^2H_8$]-glycerol/$^2H_2O$/$H_2O$ (6/3/1 v/v/v) mixture with microwave irradiation at 18.8 T and 125 K with $\nu_R$ = 24 kHz. Figure adapted from ref. [80]. Copyright 2018, American Chemical Society.

After the coherent polarization transfer from protons to another isotope, additional pulse schemes have been applied to separate the isotropic and anisotropic interactions or to probe the internuclear connectivities and the proximities. For instance, for spin-1/2 nuclei subject to large CSA, such as $^{113}Cd$ or $^{119}Sn$, techniques, such as magic-angle turning (MAT) [114] and phase adjusted spinning sideband (PASS) [159], have been used to refocus the CSA and to correlate the isotropic chemical shifts of the different sites along the indirect dimension to their MAS sideband manifolds along the direct dimension, allowing their CSA to be determined (see Figure 24). Similarly multiple-quantum (MQ) MAS (MQMAS) sequences have been used to refocus the second-order quadrupolar broadening of $^{27}Al$ and $^{17}O$ nuclei and to correlate the isotropic shifts along the indirect dimension with the MAS signals (see Figure 16) [184,225]. The DNP-enhanced $^{27}Al$ MQMAS spectra were recorded by transferring via a CP step the $^1H$ DNP-enhanced polarization to the $^{27}Al$ triple-quantum (3Q) coherences, whereas the $^{17}O$ MQMAS spectrum enhanced by $^{17}O$ direct DNP has been acquired using the conventional 3QMAS sequence using z-filter.

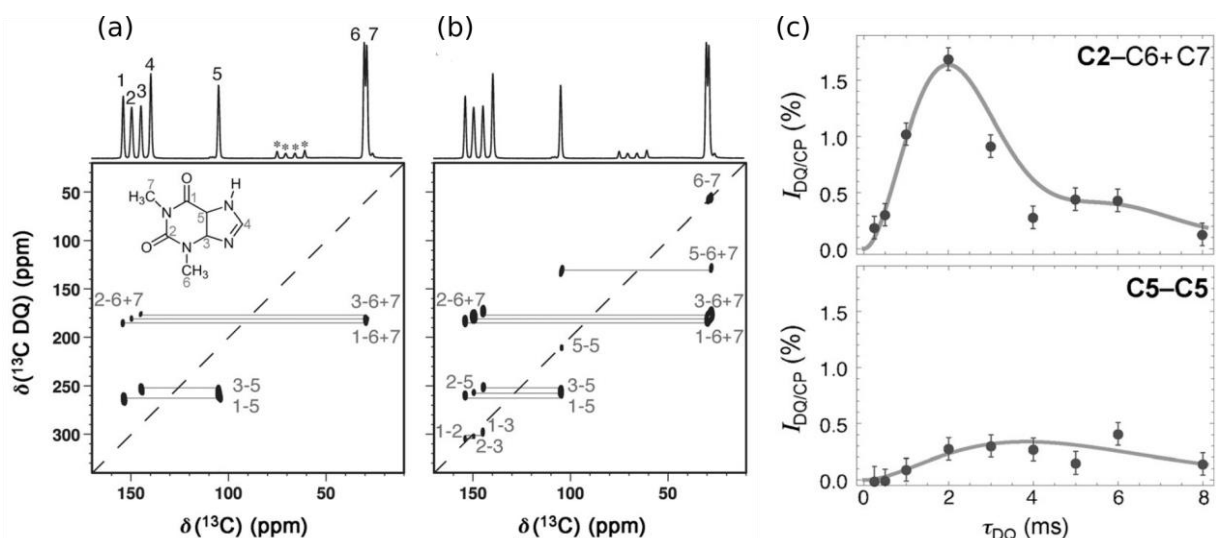

**Figure 17.** (a,b) $^{13}$C 2Q-1Q through-space homonuclear correlation 2D spectra of the isotopically unmodified anhydrous polymorph of theophylline impregnated with 66 mM TEKPol solution in 1,1,2,2-tetrabromoethane with microwave irradiation at 9.4 T and 105 K with $\nu_R$ = 8 kHz. POST-C7 scheme was used to reintroduce the $^{13}$C-$^{13}$C dipolar interactions. The length of the excitation and reconversion POST-C7 blocks was (a) 0.5 and (b) 2 ms. (c) Experimental build-up curves of the cross-peak intregated intensities (black disks) between the (top) C2 and C6+7 and (bottom) C5 signals. The best-fit curves used for the determination of the $^{13}$C-$^{13}$C distances are displayed as solid lines. Figure adapted from ref. [226]. Copyright 2015, Wiley.

Indirect DNP has also been combined with pulse sequences to probe homo- and hetero-nuclear connectivities and proximities. The sensitivity gain provided by DNP is notably useful for isotopes with low natural abundances, such as $^{13}$C, $^{15}$N or $^{29}$Si. For instance, the DNP-enhanced refocused incredible natural abundance double-quantum transfer experiment (INADEQUATE) 2D experiments have been used to observe $^{13}$C-$^{13}$C [101] and $^{29}$Si–O–$^{29}$Si [227] covalent bonds in isotopically unmodified materials, whereas the probability of these internuclear connectivities are as low as 0.01% and 0.2%, respectively (see Figure 23b).

$^{13}$C-$^{13}$C and $^{29}$Si-$^{29}$Si proximities have also been detected at natural abundance in materials using DNP-enhanced $^{13}$C [121] and $^{29}$Si [227] double-quantum (2Q)-single-quantum (1Q) through-space homonuclear correlation 2D spectra. These 2Q-1Q experiments allow for the observation of proximities between sites with close or identical isotropic chemical shifts. The $^{13}$C-$^{13}$C and $^{29}$Si-$^{29}$Si dipolar couplings have been reintroduced using symmetry-based 2Q dipolar recoupling schemes, such as POST-C7 (see Figure 17a and b) [121,227], SPC5 [171,228], SR26 [229] or S$_3$ [230]. An advantage of carrying out these experiments at natural abundance is that the dipolar truncation is significantly reduced with respect to isotopically labeled samples. For instance, it has been shown that $^{13}$C-$^{13}$C distances up to 0.7 nm can be estimated by measuring the build-up of cross-peaks from a series of through-space homonuclear correlation 2D spectra (see Figure 17) [226] or the transfer of longitudinal magnetization between two specific sites using selective pulses [230]. The second approach only requires the acquisition of a series of 1D spectra.

$^{13}$C proximities have also been probed by DNP-enhanced $^{13}$C 1Q-1Q through-space homonuclear correlation 2D spectra, which are usually more sensitive than the 2Q-1Q variants. The 1Q-1Q experiments have been acquired using dipolar assisted rotational resonance (DARR) [121,231] and CHHC [232] techniques. It has been demonstrated that the latter technique, in which the $^{13}$C-edited $^1$H polarization is transported by $^1$H-$^1$H spin diffusion, allows for the detection of long-range correlations between $^{13}$C nuclei that are close to the same network of protons [232].

$^1$H-$^1$H proximities in aminoacid microcrystals have been observed using $^1$H DNP-enhanced 2Q-1Q through-space homonuclear correlation experiments employing POST-C7 recoupling [166]. For this experiment, $^1$H spin locking has been used to filter out the signal of the solvent since in this case, solvent protons relaxed faster than those of the investigated materials. Furthermore, the resolution along the two dimensions of the 2D correlation has been improved by applying eDUMBO-122 homonuclear decoupling during the indirect evolution period, $t_1$, and the acquisition period, $t_2$.

Furthermore, $^{27}$Al-$^{27}$Al proximities have been probed using DNP-enhanced 2Q-1Q through-space homonuclear correlation experiments, in which $^{27}$Al-$^{27}$Al dipolar interactions were reintroduced under MAS by the application of BR2$_2^1$ recoupling [185]. The DNP sensitivity gain compensates for the low transfer efficiency (~5%) of this technique, which results from the intricate spin dynamics in the presence of MAS and an rf field.

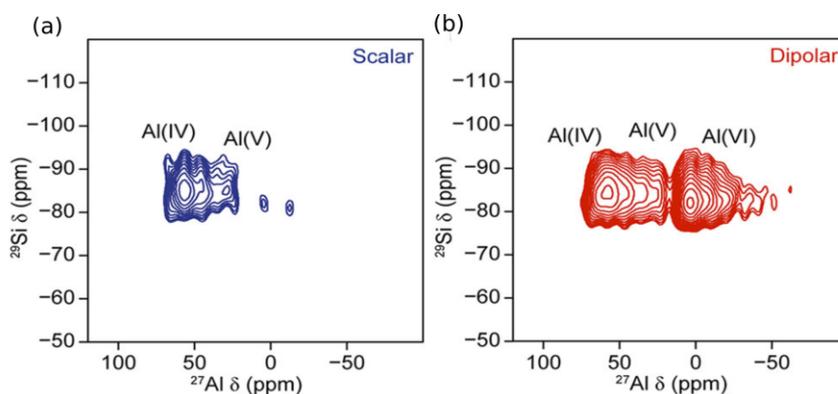

**Figure 18.** $^{27}$Al{$^{29}$Si} (a) *J*- and (b) *D*-R-INEPT 2D spectra of $^{29}$Si-enriched silicated γ-alumina (1.5% wt Si) wetted with 15 mM TEKPol solution in TCE at 9.4 T and 100 K with $\nu_R$ = 10 kHz. A $^1$H→$^{29}$Si CPMAS transfer creates the initial DNP-enhanced transverse $^{29}$Si magnetization. The spectrum of the subfigure a indicates that Si atoms are covalently connected to AlO$_4$ and AlO$_5$ sites but not to AlO$_6$ ones. Figure adapted from ref. [233]. Copyright 2017, American Chemical Society.

DNP experiments have also been used to probe heteronuclear connectivities and proximities in materials. As seen in Figure 18a, $^{27}$Al-O-$^{29}$Si connectivities at the surface of aluminosilicates have been observed using polarization transfer from $^{29}$Si to $^{27}$Al nuclei via $^2J$-coupling constants with the help of refocused insensitive nuclei enhanced by polarization transfer (*J*-R-INEPT) sequence with $^{27}$Al detection (denoted $^{27}$Al{$^{29}$Si}) [192]. One-bond $^{13}$C-$^{14}$N covalent bonds have also been observed using DNP-enhanced *J*-mediated heteronuclear multiple-quantum correlation (*J*-HMQC) sequence for the indirect detection of $^{14}$N overtone transition *via* the attached $^{13}$C nuclei [212].

$^1$H-$^{17}$O dipolar couplings have been estimated by measuring the build-up of the DNP-enhanced $^1$H→$^{17}$O PRESTO signal [222]. However, this technique lacks robustness to rf field inhomogeneity, which limits its accuracy. It has been shown that this issue can be circumvented by using the windowed-proton detected local-field (wPDLF) technique combined with $^1$H→$^{17}$O PRESTO scheme [234]. DNP-enhanced rotational-echo double resonance (REDOR) technique has been employed to estimate the dipolar couplings between pairs of spin-1/2 heteronuclei, such as $^{13}$C-$^6$Li [235], $^{13}$C-$^{15}$N [236], $^{13}$C-$^{29}$Si [237], $^{13}$C-$^{119}$Sn [238] and $^{15}$N-$^{29}$Si [236] pairs (see Figure 22). Furthermore, dipolar couplings between spin-1/2 and quadrupolar nuclei, such as $^{13}$C-$^{27}$Al, have been estimated using rotational-echo saturation pulse double-resonance (RESPDOR) technique employing simultaneous frequency and amplitude modulation (SFAM) recoupling [239]. This double-resonance $^{13}$C-$^{27}$Al experiment required the use of a frequency splitter owing to the close Larmor frequencies of these isotopes.

Proximities between spin-1/2 heteronuclei, such as $^{13}$C-$^{31}$P [210] and $^{13}$C-$^{15}$N [229], have also been probed using through-space heteronuclear correlation (HETCOR) sequences using CP transfer (see Figure 19). The sensitivity gain provided by DNP is particularly useful for detecting proximities between spin pairs having low *NA*, such as $^{13}$C-$^{15}$N with *NA* ~0.004%. Furthermore, DNP-enhanced $^{13}$C{$^{111}$Cd} through-space heteronuclear correlations have been reported using the dipolar-mediated heteronuclear multiple-quantum correlation (*D*-HMQC) sequence. This sequence avoids the use of spin lock for the $^{111}$Cd isotope subject to large CSA and so is more robust and efficient than CPMAS in this case [158]. Similarly, the *D*-HMQC sequence has been employed to observe proximities between spin-1/2 and quadrupolar nuclei, such as $^{13}$C-$^{27}$Al [239] or $^{13}$C-$^{35}$Cl [204]. $^{27}$Al-$^{29}$Si proximities have also been identified using the $^{27}$Al{$^{29}$Si} *D*-R-INEPT sequence with R$^3$ [192] or REDOR [233] schemes to reintroduce the $^{27}$Al-$^{29}$Si dipolar couplings (see Figure 18b).

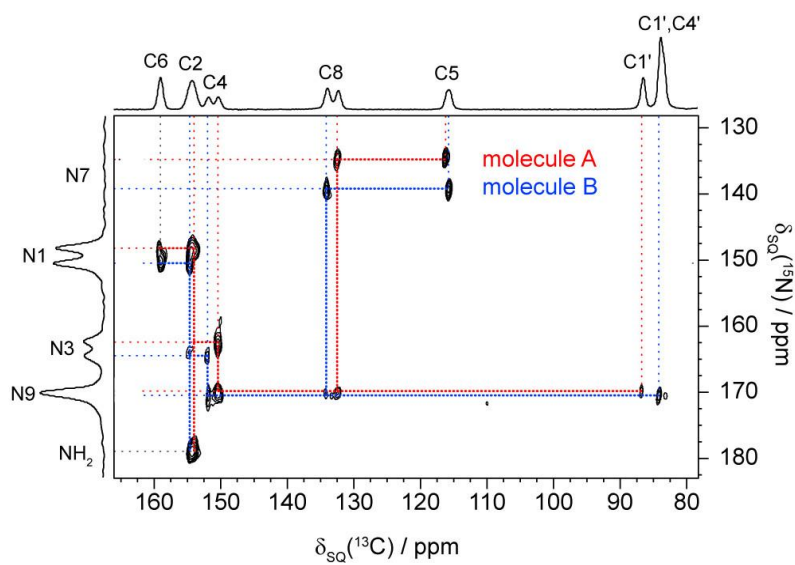

**Figure 19**. $^{13}$C{$^{15}$N} CP-HETCOR 2D spectrum of an isotopically unmodified of a self-assembled 2'-deoxyguanosine derivative with ribbon-like structure with potential application in fields ranging from medical chemistry to nanotechnology wetted with a 10 mM AMUPol solution in [$^2$H$_8$]-glycerol/$^2$H$_2$O (6/4 v/v) mixture with microwave irradiation at 9.4 T and 100 K with ν$_R$ = 12.5 kHz. Molecules A in red and B in blue denote the resonances of two distinct molecules in the asymmetric unit cell. Figure adapted from ref. [229]. Copyright 2015, American Chemical Society.

## 6. Organic materials

As seen in Figure 20, high-field MAS DNP-NMR has been used to characterize a broad range of materials, including pharmaceuticals, polymers, catalytic, energy materials, biomaterials etc. We present in this section its use for the characterization of organic materials composed of small molecules or polymers. Organic materials contain protons and so $^1$H-$^1$H spin diffusion can transport the DNP-enhanced $^1$H polarization from the surface to the bulk (see section 2.2).

### 6.1. *Pharmaceuticals and materials composed of small organic molecules*

The studies of pure and formulated active pharmaceutical ingredients using MAS DNP have been recently reviewed [65]. DNP-NMR has been used to characterize pure active pharmaceutical ingredients (API) (see Figure 12c and Figure 17) [69,163,164,204,226,240]. The sensitivity gain provided by DNP is also particularly advantageous for detecting diluted APIs (with loadings often below 10%) in polymeric excipients [104,172]. Furthermore, the sensitivity of DNP allows the problem of signal overlap in pharmaceutical formulations to be circumvented by (i) the acquisition of $^{13}$C through-bond and through-space homonuclear correlation 2D spectra [104,172] and (ii) the detection of insensitive isotopes, such as $^{15}$N [104,240] and $^{35}$Cl [204], only present in APIs. $^{1}$H-$^{15}$N through-space heteronuclear correlation 2D experiments have also been employed to probe the interactions between APIs and excipients [104]. In addition, DNP measurements allow the API domain size to be estimated using spin diffusion models (see Figure 5) [103,104]. DNP has also been used to characterize drug delivery systems, such as lipid nanoparticles containing small interfering RNA and messenger RNA [241].

As mentioned in section 3.2, PAs have been incorporated into the pharmaceutical materials by wetting with PA solution [101,104]. However, such procedure can induce polymorphic transitions and desolvation [164]. As a result, solvent-free procedures, such as spray drying and hot-melt extrusion [163,172], have been developed to overcome these issues.

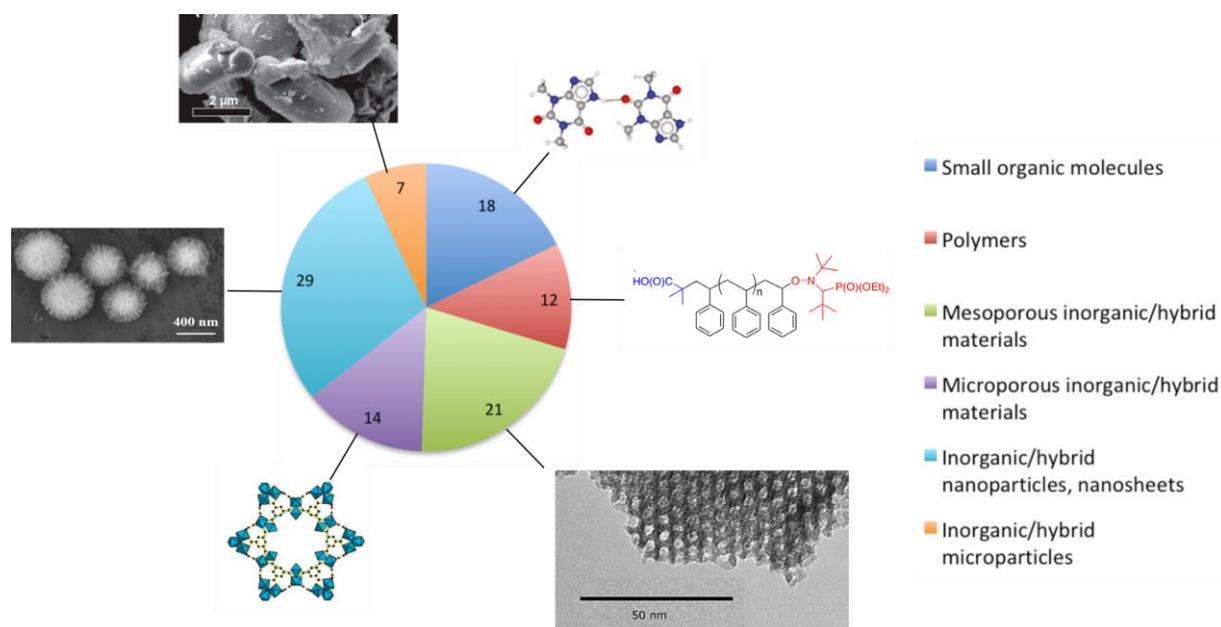

**Figure 20.** Proportions of publications reporting high-field MAS DNP NMR results for various classes of materials.

Besides APIs, DNP-NMR has been applied to study other materials made of small organic molecules (see Figure 19) [171,229,230,242]. DNP-NMR experiments have permitted the assignment of the $^{13}$C and $^{15}$N NMR signals of these self-assembled supramolecular solids by the acquisition of $^{13}$C homonuclear and $^{13}$C-$^{15}$N correlation 2D spectra at natural abundance. The observation of $^{13}$C-$^{13}$C proximities has also provided information about the formation of hydrogen bonds and π-π stacking. The low temperature, at which DNP-NMR experiments are carried out, has also been exploited to probe the distribution of Li$^+$ ions in a frozen ionic liquid salt solution [235] and to probe the evolution of crystallization processes [243]. For this latter application, the sensitivity gain provided by DNP is advantageous to characterize transient solid forms, which are only present in small quantities. It has also been shown that the transfer of DNP-enhanced $^{1}$H polarization to nearby $^{13}$C nuclei via $^{1}$H-$^{13}$C dipolar cross-relaxation can provide insights into the dynamics of frozen non-ionic surfactants [219].

### 6.2. Polymers

DNP-NMR has been applied to the characterization of organic polymers. The sensitivity provided by DNP has notably allowed the observation of the $^{13}$C NMR signals of chain-ends of living polystyrene (PS) and poly(ethylene oxide) (see Figure 21) [88,168] as well as of residues with low abundance in wild spider silks [244]. DNP high sensitivity has also been exploited to detect $^{77}$Se signal of diluted selenate ions in water remediation polymer material, despite the low natural abundance of $^{77}$Se (7.63%) and the modest amount of selenate ions (5.5 wt%) in the investigated material [245]. The high sensitivity of DNP-NMR has also been exploited to probe the surface of polymer materials, including cellulose nanocrystals or fibers [246–248] and core-shell nanoparticles [249]. DNP-enhanced $^{13}$C INADEQUATE experiments at natural abundance have also provided new insights into the atomic-level structure of biomass samples and amorphous photocatalytic polymers [221,250].

Various approaches have been employed to mix polymer samples and radicals (see section 3.2). Soluble polymers can be swollen in radical solution [153,244]. Furthermore, polymer solution casting has also been used to incorporate PAs into soluble polymers, while minimizing the amount of solvent [153,167,168]. Incipient wetness impregnation or wetting with radical solutions has also been employed for microporous organic polymers [190,191,221], nanoparticles [246–250] and thin-films [251] insoluble in the radical solution.

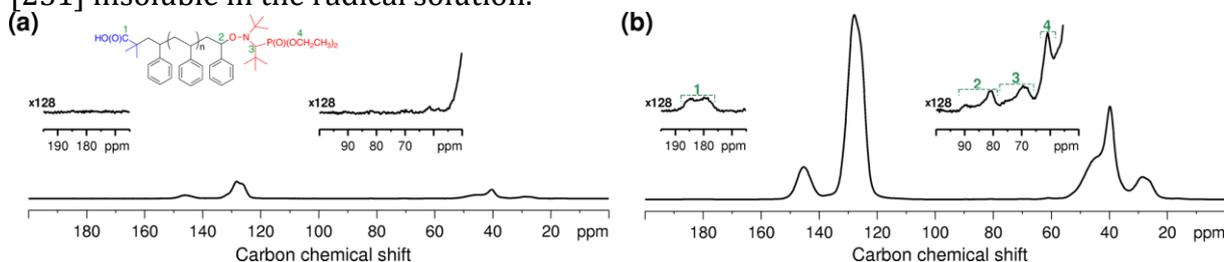

**Figure 21.** (a) Conventional $^1$H→$^{13}$C CPMAS spectrum of a living PS sample with a number average molar mass $M_n$ = 13,5 kg.mol$^{-1}$ at 285 K. (b) $^1$H→$^{13}$C CPMAS spectrum with microwave irradiation of the same polymer doped with 0.5 wt% bCTbk using the film casting procedure at 105 K. Both spectra were acquired at 9.4 T with $v_R$ = 10 kHz. Figure adapted from ref. [168]. Copyright 2013, American Chemical Society.

## 7. Inorganic and hybrid materials

High-field MAS DNP-NMR has been applied to a wide range of inorganic and hybrid materials used notably in the field of catalysis and energy. The bulk region of inorganic materials often contains no protons and in such cases, indirect DNP can be used for the selective observation of nuclei near the surface (see section 5.1). Furthermore, for most materials, the nuclei located near the surface represent a small fraction of the total number of nuclei and the sensitivity gain provided by DNP facilitates the detection of surface NMR signals. Therefore, DNP has been mainly applied to the analysis of high-surface area inorganic materials, including micro- and meso-porous materials as well as nanoparticles. Such DNP SENS experiments have previously been reviewed in ref. [51,54,57,61].

### 7.1. Mesoporous materials

Mesoporous materials are materials containing pores with diameters between 2 and 50 nm. Hence, the pores of these materials are generally accessible to the PAs, except when the pores are filled with surfactants [102] or for the materials with the smallest pore sizes when using large PAs, such as TEKPol, for which the second largest dimension is larger than the pore diameter [110].

DNP has been used to characterize mesoporous silica functionalized by organic species (see Figure 1c) [36,102,122,208,228,232,252,253] or metal complexes [134,236,254–260] as well as encapsulating metal nanoparticles [261]. The sensitivity provided by DNP is particularly useful for the detection of surface sites, notably when they are occupied by insensitive nuclei, such as $^{15}$N [236,252,253,255,257,260], $^{89}$Y [258], $^{195}$Pt [259] and $^{17}$O [222,262]. DNP-enhanced $^1$H→$^{13}$C and $^1$H→$^{29}$Si CP-HETCOR 2D spectra have also revealed interactions between the silica surface and ligands of supported complexes of Pd, Ir and Ru metals [134,254,256]. As seen in Figure 22, $^{13}$C{$^{29}$Si} and $^{29}$Si{$^{15}$N} REDOR experiments have been used to determine the 3D structure of a silica-supported Pt complex [236]. The determined structure is consistent with an interaction between the Pt metal center and the silica surface oxygen atoms. The sensitivity gain of DNP has also allowed probing the $^{13}$C-$^{13}$C and $^{29}$Si-$^{29}$Si proximities at the surface of mesoporous silica at natural abundance [228,232]. Besides mesoporous silica, DNP has been applied for other mesoporous materials, including mesoporous alumina [185], periodic mesoporous organosilicate and its post-synthesis functionalized organometallic derivatives [263] as well as $Ca_3Al_2O_6$-stabilized CaO [197], which is a $CO_2$ sorbent.

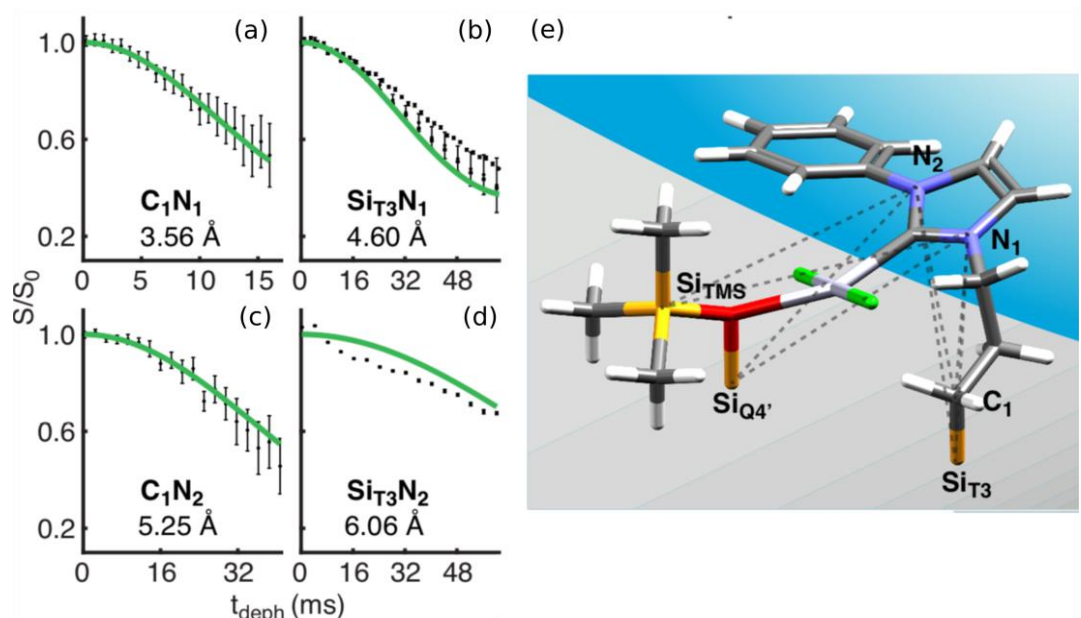

**Figure 22.** (a,c) Experimental $^{13}$C{$^{29}$Si} and (b,d) $^{29}$Si{$^{15}$N} REDOR signal (black dots) as function of recoupling time of mesoporous silica immobilizing Pt complexes selectively labeled with $^{15}$N on sites (a,b) $N_1$ and (c,d) $N_2$ impregnated with 16 mM solution of TEKPol2 in TCE with microwave irradiation at 9.4 T and 110 K with $\nu_R$ = 8 kHz. The best fit curves are shown as solid green lines. (e) Best-fit 3D structure of the surface species. Figure adapted from ref. [236]. Copyright 2017, American Chemical Society.

### 7.2. Microporous materials

Microporous materials contain pores with diameters less than 2 nm. Hence, the pores of these materials may not be accessible to PAs and the polarization is transported into the particles via spin diffusion among the protons of the frozen solvent contained within the micropores and of the investigated material itself [111,113–115].

For non-functionalized MOFs, small nitroxide biradicals, such as TOTAPOL or bTbk, can slowly diffuse into the pores [111,239]. Such diffusion has been monitored by EPR spectroscopy and the estimate of $^{13}$C-$^{27}$Al dipolar couplings in aluminium-based MOF has demonstrated that the impregnation with a radical solution and the freezing of the sample down to 100 K do not alter the particular MOF's structure [239]. For bulkier PAs, such as TEKPol, or functionalized MOFs, the PAs cannot penetrate the pores and reside on or near the surface of the crystallites [111–113,264]. DNP-enhanced $^1$H polarization is transported within the microcrystal by spin diffusion among protons of the linker and the solvent. However, such transport leads to losses of polarization. Hence, the DNP enhancements for MOFs are in general significantly lower than those achieved for mesoporous materials. Nevertheless, the gain in sensitivity provided by DNP has enabled the detection of insensitive isotopes, such as $^{15}$N in natural abundance [111,113,264] or $^{195}$Pt [112], in MOFs.

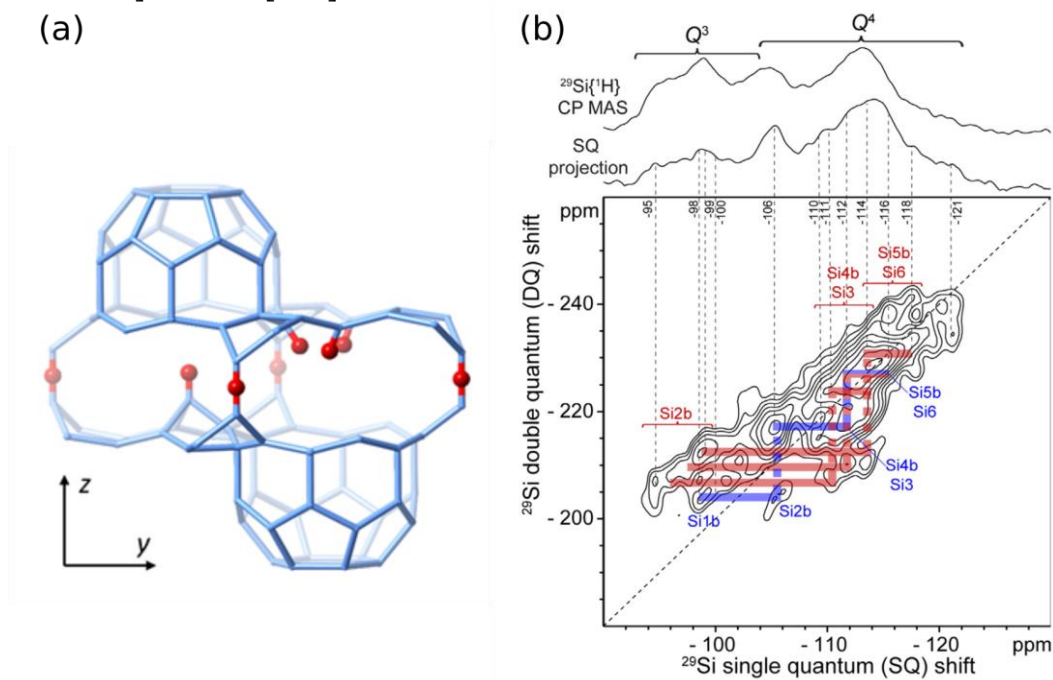

**Figure 23.** (a) Schematic structure of the interlayer region of the calcined form of the zeolite SSZ-70. The terminal and interlayer-bridging O atoms are shown in red, whereas the others have been omitted. (b) DNP-enhanced $^{29}$Si refocused INADEQUATE 2D spectrum of isotopically unmodified SSZ-70 zeolite at 9.4 T and ∼100 K with $\nu_R$ = 8 kHz. The DNP-enhanced $^1$H→$^{29}$Si CPMAS 1D spectrum acquired under the same conditions is shown along the 1Q dimension. The blue and red lines indicate the connectivity paths to isolated silanol (on the left of subfigure a) and nested silanol groups (on the right of subfigure b), respectively. Figure adapted from ref. [265]. Copyright 2017, American Chemical Society.

DNP has also been employed for the characterization of zeolite frameworks [114–116,194] and the organic species in its micropores [117,266,267]. Similar to MOFs, only small nitroxide biradicals, such as TOTAPOL and bTbk, can penetrate into the pores of zeolites [115]. The bulkier radicals, such as bCTbk and TEKPol, remain near the surface of the particles and the transport of DNP-enhanced $^1$H polarization to the internal sites rely exclusively on $^1$H-$^1$H spin diffusion [114–116,194], which limits the achievable DNP enhancements. Recently, it has been shown that the introduction in zeolites of hierarchical pores, i.e. of macropores with diameter greater than 50 nm and mesopores connected to the micropores, improves the DNP enhancement [117].

The DNP sensitivity gain has been exploited to detect $^{119}$Sn nuclei exhibiting large CSA in zeolites loaded with only a few percent of tin [114–116,194] and to acquire $^{29}$Si refocused INADEQUATE 2D spectra at natural abundance (see Figure 23) [265]. Furthermore, $^{13}$C refocused INADEQUATE and $^{29}$Si{$^{13}$C} REDOR experiments combined with DNP and $^{13}$C isotopic enrichment have been applied to probe the structure of carbocations on zeolites and their interactions with the frameworks [117].

### 7.3. Nanoparticles and nanosheets

Nanoparticles are particles between 1 and 100 nm in size, whereas nanosheets are two-dimensional nanostructures with a thickness in a scale ranging from 1 to 100 nm. The surface of these nanostructures is generally accessible to the PAs, except when the nanostructures aggregate [118,158]. Methods to prevent this aggregation have been discussed in section 3.2 . DNP has been applied to characterize the surface of nanoparticles used for catalysis [124,184,192,238,268], biomaterials [106,107], cements [269], polymer fillers [85,210] and optoelectronics devices [158,159]. The investigated nanoparticles included functionalized silica [193,227,238,270], alumina [80,124,184,196,225,231,271–273], silica alumina [192,233,234,274], ceria [195,275], sulfated zirconia [276], calcium silicate hydrates [269], phosphates [106,107,210], partially oxidized Sn nanoparticles [151] and crystalline semiconductors, such as CdSe, CdS or InP, in the form of nanoparticles (also called quantum dots) [158,159]. DNP has also been applied to investigate calcium silicate hydrates [269] and crystalline semiconductors, such as CdSe and CdS, in the form of nanosheets (also called nanoplatelets) (see Figure 24) [159].

The sensitivity gain provided by DNP has been useful to detect insensitive isotopes with low natural abundance, such as $^{15}$N [196,268], $^{43}$Ca [107] or $^{17}$O [225,234], or low γ ratio, such as $^{89}$Y [270]. The high sensitivity of DNP is also beneficial to observe spin-1/2 nuclei subject to large CSA, such as $^{119}$Sn and $^{113}$Cd [151,158,159,238]. For instance, PASS technique has allowed discerning the isotropic chemical shifts and estimating the CSA of the various $^{113}$Cd sites in CdSe and core/crown CdSe/CdS nanoplatelets (see Figure 24) [159]. It has been shown that $^{13}$C-$^{13}$C and $^{29}$Si-$^{29}$Si connectivities and proximities in nanoparticles can be probed using DNP-enhanced through-space and through-bond 2Q-1Q homonuclear correlation experiments at natural abundance [106,227,269]. Nevertheless, the determination of the structure of reaction intermediates and products at the surface of catalysts has required the use of $^{13}$C isotopic enrichment, in spite of the sensitivity gain provided by DNP [193,231,271]. DNP has also been exploited to probe heteronuclear proximities involving a low abundant isotope, such as $^{13}$C-$^{31}$P [106,210] and $^{29}$Si-$^{27}$Al (see Figure 18) [192,233], in nanoparticles.

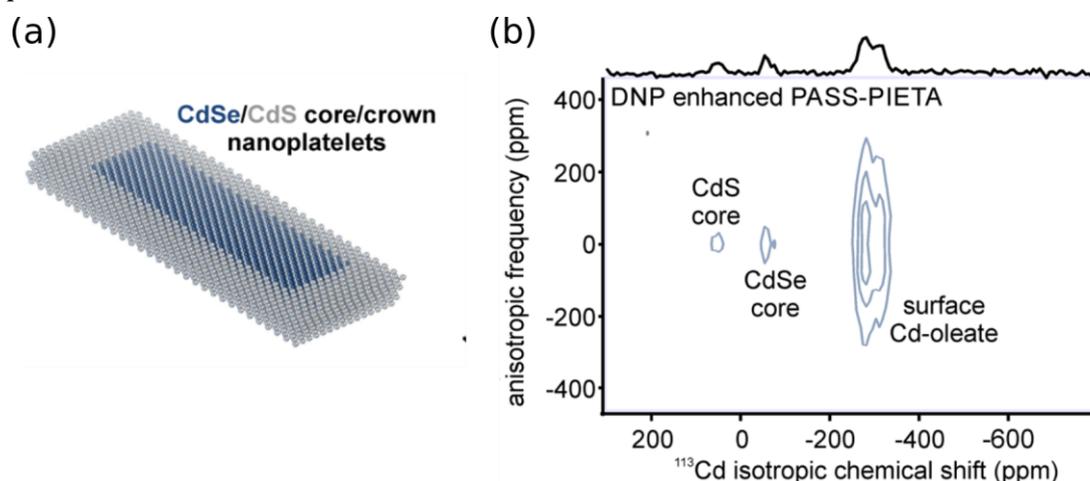

**Figure 24.** (a) Atomic model of Cd-terminated core/crown CdSe/CdS nanoplatelets. (b) $^{113}$Cd PASS 2D spectrum enhanced by phase-incremented echo-train acquisition (PIETA) of the oleate-capped core/crown CdSe/CdS nanoplatelets sol in TCE mixed with 50 mM TEKPOL solution in TCE yielding a final radical concentration of 16 mM and impregnating a mesoporous silica at 14.1 T and 100 K with $v_R$ = 10 kHz. Four Cd species can be resolved and assigned to cadmium in the CdS core (57 ppm), the CdSe core (−61 ppm), and to Cd-oleate at the surface of CdS (−281 ppm) and CdSe (−310 ppm). Figure adapted from ref. [159]. Copyright 2018, American Chemical Society.

### 7.4. Microparticles

Microparticles are particles between 0.1 and 100 μm in size. Hence, the nuclei in the microparticle core are distant from the PAs, except when the core is doped with PAs [71–73]. Nevertheless, for proton-containing microparticles, the $^1$H-$^1$H spin diffusion can transport the DNP-enhanced $^1$H polarization (see section 2.2) and significant DNP enhancements have been reported for several proton-containing inorganic or hybrid microparticles, including BaZrO$_3$-based proton conductors with diameter up to 45 μm [109], thermally treated ammonia borane, a promising material for hydrogen storage, with diameter up to 8 μm [108], riverine organic matter [277], solid electrolyte for Li-ion cells [162] and lead-based pigment and painting [207]. The DNP sensitivity enhancement has notably been used to detect the signals of insensitive nuclei, such as $^{15}$N [108] and $^{89}$Y [109], with low gyromagnetic ratio and low natural abundance or of isotopes subject to large CSA [207] in microparticles. DNP has also been exploited to enhance the $^{13}$C signals of the outer layers of solid electrolyte interphase [162] and to observe $^{13}$C-$^{27}$Al proximities in natural abundance in riverine organic matter containing 1.4% of aluminium atoms [277].

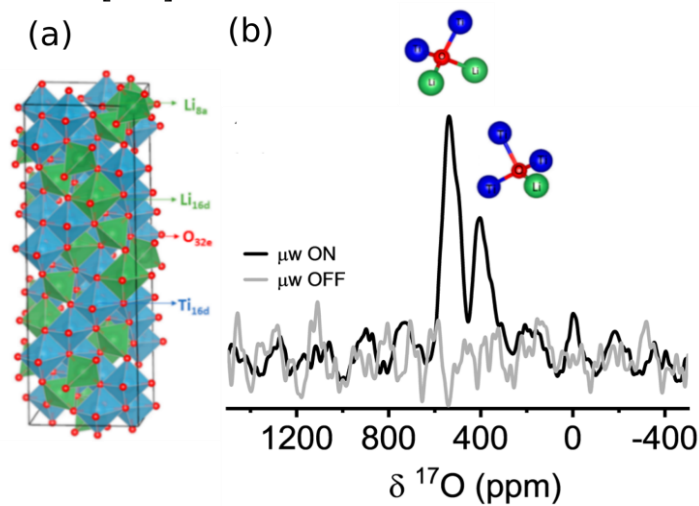

**Figure 25.** (a) Supercell for one of the two configurations of Li$_4$Ti$_5$O$_{12}$ exhibiting the lowest energy. (b) $^{17}$O NMR spectrum of Li$_4$Ti$_5$O$_{12}$ doped Mn(II) ions at 0.5% molar fraction with (black solid line) and without (grey solid line) microwave irradiation at 9.4 T and around 100 K with $v_R$ = 10 kHz. Figure adapted from ref. [73]. Copyright 2019, American Chemical Society.

Furthermore, in the case of inorganic microparticles, which do not contain proton, indirect DNP has been used to detect the signals of nuclei near the particle surface. A challenge is the low surface area of microparticles, which leads to a small fraction of nuclei in the vicinity of surface. Nevertheless, DNP has been successfully employed to probe the molecular-level interactions between the surface sites of triclinic Ca$_3$SiO$_5$ microparticles, a major component of aluminosilicate cements, with a surface area of 1 m$^2$.g$^{-1}$ and hydration inhibitors (sucrose or methylenephosphonic acid) at a loading of 0.1% by weight of solids in the presence of water [237]. More recently, sensitivity gain provided by DNP has allowed the acquisition of $^{31}$P signals of oligonucleotide deposited onto silicate and glass wafers with surface areas on the order of ca. 0.01 m$^2$.g$^{-1}$ [278].

DNP has also been applied to enhance the NMR signals of the core region of inorganic microparticles. Such enhancement is possible, when the core of the microparticles contains PAs [71–73]. DNP has particularly been used to detect the distinct $^{17}$O sites of crystalline inorganic microparticles doped with Mn(II) ions, including battery anode materials Li$_4$Ti$_5$O$_{12}$ and Li$_2$ZnTi$_3$O$_8$, as well as the materials NaCaPO$_4$ and MgAl$_2$O$_4$ (see Figure 25) [73].

## 8. Conclusion

We have presented here the recent developments in high-field MAS DNP-NMR of materials. The development of novel DNP-NMR instruments, the design of tailored PAs, the optimization of sample preparation, and the better understanding of DNP mechanisms under MAS have all significantly improved the sensitivity of this technique and broadened its application fields. Currently MAS DNP has been applied to a wide range of materials, including organic, hybrid and inorganic materials with applications to fields, such as catalysis, polymers, health, optoelectronics, etc. The DNP sensitivity enhancement has notably provided unique insights for the characterization of pharmaceutical formulations and the surface of inorganic particles and mesoporous materials.

Nevertheless, the polarization enhancements produced by DNP with respect to thermal equilibrium are still much lower than the maximum theoretical enhancement for continuous-wave DNP, which is given by the ratio of the gyromagnetic ratios of the unpaired electron and the polarized isotope, $\gamma(S)/\gamma(I)$. For instance, for a frozen solution of 20 mM [$^{13}$C]-urea and 10 mM AsymPolPOK in glycerol/$^2$H$_2$O/H$_2$O (60/30/10 v/v/v) at 9.4 T and 105 K with $\nu_R$ = 10 kHz, DNP yields a 83-fold enhancement of the nuclear polarization, which is still 8 times smaller than $\gamma(S)/\gamma(^1H) \approx 660$ [80]. Furthermore, the DNP enhancement decreases at higher $B_0$ field. For a frozen solution of 1 M [$^{13}$C]-urea and 10 mM TEMTriPol-1 in glycerol/$^2$H$_2$O/H$_2$O (60/30/10 v/v/v) at 18.8 T and 125 K with $\nu_R$ = 10 kHz, the polarization enhancement is only equal to 51 [138].

In the future, promising approaches to further improve the absolute sensitivity of MAS DNP include the developments of (i) optimized PAs, notably for high magnetic fields [81,139], (ii) MAS at temperatures below 77 K with closed-loop cold helium system [19,62,179,199–201,279], (iii) improved MAS DNP-NMR probes with better coupling between the incident microwave and the sample space [139,178], (iv) frequency-swept DNP microwave sources, which can improve electron decoupling, and hence, reduce linewidth and enhance signal intensity [19,280], and (v) DNP microwave sources capable of phase coherent pulses with a length of nanoseconds to microseconds [281] combined with pulsed DNP techniques [282–286]. These latter DNP techniques may improve the sensitivity at high fields since their efficiency is a priori independent of the $B_0$ field. Furthermore, they do not require biradical PAs contrary to CE mechanism and may be applicable at higher temperatures than continuous wave DNP.

DNP experiments at higher MAS frequencies are also desirable for (i) the acquisition of high-resolution $^1$H NMR spectra and (ii) the use of indirect detection via protons [16]. Higher MAS frequencies can be achieved by the development of (i) DNP-NMR probes for rotors with outer diameter smaller than 1.3 mm and (ii) spinning systems using helium gas (see section 4.5) [19,62,179,199–201,279]. The small diameter rotors will also be useful for the study of volume-limited samples [251] and experiments with high rf field requirements, such as MQMAS.

The advent of lower-cost DNP-NMR instrumentations, including more economic microwave sources, such as klystrons [202,279] or solid-state sources [199,287,288], closed-loop cryogenic system for MAS of the sample [179,201], and DNP-NMR probes for narrow-bore NMR magnets, will further increase the availability of this technique and the number of users and potential applications.

The DNP-NMR study of materials also requires the optimization of the PAs and the sample preparation, notably to limit the reaction between the investigated material and the PAs [110,118,135], to avoid the use of solvents [121,168,172] or to enhance the signals of the bulk region of inorganic materials [71,73].

Further improvements in the sensitivity of DNP-NMR will open new avenues for the characterization of materials. It will particularly allow for the observation of diluted sites corresponding to amounts of matter in the nanomole-scale range or below. In particular, these amounts are those of the surface sites of materials with low surface area, such as silicon wafers used in microelectronics with surface area in the order of $10^{-3}$ mm$^2$.g$^{-1}$ [278]. High DNP enhancements at very high magnetic fields, $B_0 > 17$ T, will notably offer new possibilities for the detection of quadrupolar nuclei [80,95] and especially those with low gyromagnetic ratio, such as $^{25}$Mg, $^{33}$S, $^{35}$Cl, $^{67}$Zn, $^{95}$Mo… The above developments will expand the scope of the application of MAS DNP-NMR.

**Acknowledgements**

The authors would like to thank the anonymous reviewers as well as Gaël De Paëpe, Franck Engelke, Takeshi Kobayashi, Michal Leskes, Frédéric Mentink-Vigier and Giulia Mollica for their comments that greatly contributed to improve the quality of the manuscript. Chevreul Institute (FR 2638), Ministère de l'Enseignement Supérieur, de la Recherche et de l'Innovation, Hauts-de-France Region and FEDER are acknowledged for supporting and funding partially this work. Financial support from the IR-RMN-THC FR-3050 CNRS for conducting the research is gratefully acknowledged. Authors also thank contracts ANR-17-ERC2-0022 (EOS) and ANR-18-CE08-0015-01 (ThinGlass). This project has received funding from the European Union's Horizon 2020 research and innovation program under grant agreement No 731019 (EUSMI). OL acknowledge financial support from Institut Universitaire de France (IUF).

methods, Magn. Reson. Chem. 53 (2015) 927–939. doi:10.1002/mrc.4290.

[11]  Z. Gan, P. Gor'kov, T.A. Cross, A. Samoson, D. Massiot, Seeking higher resolution and sensitivity for NMR of quadrupolar nuclei at ultrahigh magnetic fields., J Am Chem Soc. 124 (2002) 5634–5635. doi:10.1021/ja025849p.

[12]  K. Hashi, S. Ohki, S. Matsumoto, G. Nishijima, A. Goto, K. Deguchi, K. Yamada, T. Noguchi, S. Sakai, M. Takahashi, Y. Yanagisawa, S. Iguchi, T. Yamazaki, H. Maeda, R. Tanaka, T. Nemoto, H. Suematsu, T. Miki, K. Saito, T. Shimizu, Achievement of 1020MHz NMR, J. Magn. Reson. 256 (2015) 30–33. doi:10.1016/j.jmr.2015.04.009.

[13]  Z. Gan, I. Hung, X. Wang, J. Paulino, G. Wu, I.M. Litvak, P.L. Gor'kov, W.W. Brey, P. Lendi, J.L. Schiano, M.D. Bird, I.R. Dixon, J. Toth, G.S. Boebinger, T.A. Cross, NMR spectroscopy up to 35.2T using a series-connected hybrid magnet, J. Magn. Reson. 284 (2017) 125–136. doi:10.1016/j.jmr.2017.08.007.

[14]  S. Hartmann, E. Hahn, Nuclear Double Resonance in the Rotating Frame, Phys. Rev. 128 (1962) 2042–2053. doi:10.1103/PhysRev.128.2042.

[15]  A. Pines, M.G. Gibby, J.S. Waugh, Proton‐enhanced NMR of dilute spins in solids, J. Chem. Phys. 59 (1973) 569–590. doi:10.1063/1.1680061.

[16]  Y. Ishii, R. Tycko, Sensitivity Enhancement in Solid State $^{15}$N NMR by Indirect Detection with High-Speed Magic Angle Spinning, J. Magn. Reson. 142 (2000) 199–204. doi:10.1006/jmre.1999.1976.

[17]  P.C. Myhre, G.G. Webb, C.S. Yannoni, Magic angle spinning nuclear magnetic resonance near liquid-helium temperatures. Variable-temperature CPMAS studies of $C_4H_7^+$ to 5 K, J. Am. Chem. Soc. 112 (1990) 8992–8994. doi:10.1021/ja00180a061.

[18]  M. Concistre, O.G. Johannessen, E. Carignani, M. Geppi, M.H. Levitt, Magic-angle spinning NMR of cold samples, Acc. Chem. Res. 46 (2013) 1914–1922. doi:10.1021/ar300323c.

[19]  E.L. Sesti, E.P. Saliba, N. Alaniva, A.B. Barnes, Electron decoupling with cross polarization and dynamic nuclear polarization below 6 K, J. Magn. Reson. 295 (2018) 1–5. doi:10.1016/j.jmr.2018.07.016.

[20]  R.A. Wind, F.E. Anthonio, M.J. Duijvestijn, J. Smidt, J. Trommel, G.M.C. de Vette, Experimental Setup for Enhanced $^{13}$C NMR Spectroscopy in Solids Using Dynamic Nuclear Polarization, J Magn Reson. 52 (1983) 424–434. doi:10.1016/0022-2364(83)90168-3.

[21]  D.A. Hall, D.C. Maus, G.J. Gerfen, S.J. Inati, L.R. Becerra, F.W. Dahlquist, R.G. Griffin, Polarization-Enhanced NMR Spectroscopy of Biomolecules in Frozen Solution, Science. 276 (1997) 930–932. doi:10.1126/science.276.5314.930.

[22]  R.A. Wind, Dynamic Nuclear Polarization and High-Resolution NMR of Solids, in: R.K. Harris, R.E. Wasylishen (Eds.), Encycl. Magn. Reson., John Wiley & Sons, Ltd, Chichester, UK, 2007.

[23]  V.K. Michaelis, T.-C. Ong, M.K. Kiesewetter, D.K. Frantz, J.J. Walish, E. Ravera, C. Luchinat, T.M. Swager, R.G. Griffin, Topical Developments in High-Field Dynamic Nuclear Polarization, Isr. J. Chem. 54 (2014) 207–221. doi:10.1002/ijch.201300126.

[24]  A.S. Lilly Thankamony, J.J. Wittmann, M. Kaushik, B. Corzilius, Dynamic nuclear polarization for sensitivity enhancement in modern solid-state NMR, Prog. Nucl. Magn. Reson. Spectrosc. 102–103 (2017) 120–195. doi:10.1016/j.pnmrs.2017.06.002.

[25]  D. Raftery, E. MacNamara, G. Fisher, C.V. Rice, J. Smith, Optical Pumping and Magic Angle Spinning:  Sensitivity and Resolution Enhancement for Surface NMR Obtained with Laser-Polarized Xenon, J. Am. Chem. Soc. 119 (1997) 8746–8747. doi:10.1021/ja972035d.

[26]  E. Brunner, R. Seydoux, M. Haake, A. Pines, J.A. Reimer, Surface NMR Using Laser-
42